\documentclass[twocolumn]{revtex4-1}

\usepackage{amsmath,amssymb}
\usepackage{graphicx}
\usepackage{dcolumn}
\usepackage{bm}
\usepackage{multirow}

\usepackage{natbib}

\def \bsym {\boldsymbol}
\def \n {\nonumber\\}

\begin{document}
\title{Foldy-Wouthuysen transformation for gapped Dirac fermions in two-dimensional
semiconducting materials and valley excitons under external fields}
\author{Yao-Wen Chang}
\email{yaowen920@gmail.com}
\author{Yia-Chung Chang}
\email{yiachang@gate.sinica.edu.tw}
\affiliation{Research Center for Applied Sciences, Academia Sinica, Taipei 11529, Taiwan}
\begin{abstract}
  In this work, we provide a detailed derivation of Foldy-Wouthuysen (FW) transformation
  for two-dimensional (2D) gapped Dirac fermions under external fields and apply the
  formalism to study valley excitons in 2D semiconducting materials. Similar to
  relativistic quantum few-body problem, the gapped Dirac equation can be transformed into
  a Schr\"{o}dinger equation with "relativistic" correction terms. In this 2D materials
  system, the correction terms can be interpreted as the Berry-curvature effect. The
  Hamiltonian for a valley exciton in external fields can be written based on
  the FW transformed Dirac Hamiltonian. Various valley-dependent effects on excitons, such
  as fine-structure splittings of exciton energy levels, valley-selected exciton transitions, and exciton
  valley Zeeman effect are discussed within this framework.
\end{abstract}

\maketitle

\section{Introduction}

Two-dimensional (2D) semiconducting materials such as transition metal dichalcogenides
(TMDCs) are atomic thin semiconductors known for their potential for future applications in
electronic and photonic devices\cite{wang2012electronics, xia2014two, mak2016photonics}.
The valley-dependent electronic structure of 2D materials provides a new degree of freedom
to manipulate, and the unique topology of band structure  introduces an intriguing
Berry-curvature effect on their physical properties\cite{schaibley2016valleytronics,
vitale2018valleytronics, mak2018light, zhao2021valley}. The lower dimensionality also
induces strong electron correlations and thus excitonic effect becomes important. A valley
exciton is a Wannier-like exciton whose properties are decided by the additional valley
degree of freedom and affected by the band-structure geometry\cite{cai2013magnetic,
berkelbach2015bright, yu2015valley, berkelbach2017optical, durnev2018excitons,
li2020fine}. Various physical phenomena related to valley excitons, such as
Berry-curvature induced exciton energy-level splitting\cite{srivastava2015signatures,
zhou2015berry, trushin2018model, van2019spectrum, yong2019valley}, valley-selected optical
transition\cite{berkelbach2015bright, yu2015valley, gong2017optical, zhang2018optical},
exciton valley Hall effect\cite{onga2017exciton, glazov2020skew}, and exciton valley
Zeeman effect\cite{van2018strong, bragancca2019magnetic, catarina2019optical,
koperski2018orbital, YCPRL} have been observed experimentally and discussed theoretically.
While there are a lot of theoretical works using different methods to study different
issues of valley excitons, the connection among different theories and interpretations is
not manifest.

One of theoretical issues to study valley excitons is to include external-field
interactions to the exciton model. For a Wannier exciton, which consists of an
electron and a hole bound by the Coulomb interaction, the system is
similar to a hydrogen atom in external fields. The electric-field interaction
can be included by the dipole-field interaction and the magnetic-field interaction can be
included as the vector fields in the kinetic momentums of the electron and hole. For
the valley excitons in 2D materials, the exciton Hamiltonian is constructed by the
Bloch-electron wavefunctions solved from diagonalizing a Dirac Hamiltonian or a few-band
tight-binding Hamiltonian\cite{berkelbach2015bright, berkelbach2017optical}. However, the
Dirac Hamiltonian or tight-binding Hamiltonian with including external-field interactions
is much more difficult to solve, such that the exciton Hamiltonian under external field
also becomes more difficult to derive. To simplify the problem,
approximations are necessary. In fact, a similar problem had been intensively studied  in the fields of relativistic few-body
physics\cite{bethe2012quantum, reiher2014relativistic}. A relativistic particle which is
described by a three-dimensional Dirac fermion under external fields can be reduced to a
nonrelativistic particle with relativistic corrections in low-energy limit by the
Foldy-Wouthuysen (FW) transformation\cite{reiher2014relativistic, foldy1950dirac,
eriksen1958foldy, silenko2008foldy, silenko2016general}. A relativistic few-body problem,
such as a real hydrogen atom in external fields, can be studied by solving the
nonrelativistic Hamiltonian with including the relativistic
corrections\cite{close1970relativistic, krajcik1974relativistic, anthony1994relativistic}.
The approximation scheme can achieve extreme accuracy in prediction of energy spectra and
fine structures. Following the same strategy, the 2D Dirac fermions can also be reduced to nonrelativistic form
by the FW transformation, and the valley-exciton Hamiltonian can be approximated as the
Wannier-exciton Hamiltonian with band-geometry corrections. The valley excitons in
external fields can then be studied by the corrected exciton Hamiltonian.

Semiclassical methods have been used to study the band-structure correction to the
properties of excitons\cite{xiao2005berry, yao2008valley, yao2008berry, chang2008berry,
gradhand2012first}. The Berry curvature effect on excitons in 2D materials has been
studied theoretically\cite{cai2013magnetic, srivastava2015signatures, zhou2015berry,
gong2017optical, trushin2018model, zhang2018optical, van2019spectrum} and observed
experimentally\cite{yong2019valley}. The Berry-curvature effect causes energy-level
splitting\cite{srivastava2015signatures, zhou2015berry, trushin2018model, van2019spectrum,
yong2019valley} and anomalous selection rule\cite{gong2017optical, zhang2018optical} for
valley excitons with nonzero angular momentum. It has been shown that the exciton
Hamiltonian can be derived from a FW transformation of 2D gapped Dirac
fermions\cite{zhou2015berry, trushin2018model}. However, the joint effect of the
Berry-curvature effect and external fields has yet to be discussed. We intend to fill
this gap by extending the derivation of FW transformation to incorporate both effects for 2D gapped Dirac
fermions in external fields.

In the present work, the Bloch electron in 2D materials under external fields is described
by a 2D gapped Dirac model, and the FW transformation is used to derive the electron-hole
representation of the Dirac fermion. The FW transformed single-particle Hamiltonian,
two-particle interaction, and interband transition are derived. Hamiltonians for valley
excitons in external fields, including an in-plane electric field, an in-plane
electromagnetic field, and an out-of-plane magnetic field can be written. The formalism
is applied to the study of physical properties of valley excitons. Within this theoretical
framework, several known valley-dependent excitonic effects including exciton energy-level
splittings, valley-selected exciton transitions, and exciton valley Zeeman effect are
studied and discussed.

The remainder of the article is organized as follows. In Sec.~\ref{sec:dirac_fermion}, we review
the 2D gapped Dirac fermion model for the band structure of Bloch electrons with a
screened Coulomb interaction. The electromagnetic interaction, electron-hole
representation, and exciton states based on the Dirac fermion model are also defined here.
In Sec.~\ref{sec:fw_transformation}, the FW transformation for the 2D gapped Dirac fermion
model is introduced. The FW transformed single-particle Hamiltonian, two-particle
interaction, and interband transition are derived. In Sec.~\ref{sec:valley_exciton}, the
formula derived from the FW transformation are applied to valley excitons in 2D materials. The
Hamiltonians for valley excitons in an in-plane electric field, an in-plane
electromagnetic field, and an out-of-plane magnetic field are derived, and valley-dependent
phenomena are studied. Finally, the summary and some perspectives of
this theoretical framework are given in Sec.~\ref{sec:summary}. In
Appendix~\ref{sec:variation}, the variational method used to solve exciton eigenenergies and
exciton wavefunctions is presented in detail.

\section{2D gapped Dirac fermion\label{sec:dirac_fermion}}

Before introducing the FW transformation, we first review the 2D gapped Dirac model and
derive the exciton model based on the formulation of the Dirac model. In
Sec.~\ref{sec:dirac_model}, the many-body Dirac Hamiltonian is introduced in the
second quantization formalism. In Sec.~\ref{sec:electromagnetic}, the electromagnetic-field
interaction in the Dirac model and the formula for calculating optical spectra are given.
In Sec.~\ref{sec:e_h_representation}, the electron-hole representation of the many-body
Dirac Hamiltonian is introduced and the exciton state is written in this
representation.

\subsection{Many-body electronic Hamiltonian\label{sec:dirac_model}}

The many-body electronic Hamiltonian for 2D gapped Dirac fermions can be written as
\cite{berkelbach2015bright, berkelbach2017optical}
\begin{eqnarray}
  \hat{\mathcal{H}}
  &=&
  \sum_{\tau}\int\hat{\psi}^{\dagger}_{\tau}(\mathbf{r})
  \mathtt{h}_{\tau}(\mathbf{r})\hat{\psi}_{\tau}(\mathbf{r})\text{d}^2r\n
  &&+\frac{1}{2}\int{V}({r}_{12})
  \hat{\rho}(\mathbf{r}_1)\hat{\rho}(\mathbf{r}_2)\text{d}^2r_1\text{d}^2r_2,
\end{eqnarray}
where $\hat{\psi}^{\dagger}_{\tau}(\mathbf{r})$, $\hat{\psi}_{\tau}(\mathbf{r})$ are the
fermion field operators with $\tau$ being the valley index, $\hat{\rho}(\mathbf{r}) =
\sum_{\tau} \hat{\psi}^{\dagger}_{\tau}(\mathbf{r})\hat{\psi}_{\tau}(\mathbf{r})$ is the
density operator, $\mathtt{h}_{\tau}(\mathbf{r})$ is the single-particle Hamiltonian, and
${V}({r}_{12})$ is a 2D screened Coulomb potential with
${r}_{12}=|\mathbf{r}_1-\mathbf{r}_2|$. The single-particle Hamiltonian is given by
\begin{eqnarray}
  \mathtt{h}_{\tau}(\mathbf{r})
  &=&
  v_{\text{F}}\bsym{\pi}\cdot{\bsym{\alpha}_{\tau}}+mv^2_{\text{F}}{\beta}
  -q\mathbf{F}\cdot\mathbf{r}\n
  &=&
  \begin{pmatrix}
    mv^2_{\text{F}}-q\mathbf{F}\cdot\mathbf{r}
    & v_{\text{F}}\left(\tau\pi^{x}-\mathtt{i}\pi^{y}\right)\\
    v_{\text{F}}\left(\tau\pi^{x}+\mathtt{i}\pi^{y}\right)
    & -mv^2_{\text{F}}-q\mathbf{F}\cdot\mathbf{r}
  \end{pmatrix}
\end{eqnarray}
with $\bsym{\alpha}_{\tau}=\tau\sigma_x\mathbf{e}_{x}+\sigma_y\mathbf{e}_{y}$ and
$\beta=\sigma_{z}$, and $\sigma_{x}$, $\sigma_{y}$, $\sigma_{z}$ being Pauli matrices,
$\mathbf{F}$ an in-plane electric field, $\bsym{\pi}$ the kinetic momentum, $q$ the
charge, $v_{\text{F}}$ the Fermi velocity, and $m$ the effective mass of the particle. The
effective mass can be connected to a band-gap energy by $\Delta_0=2mv^2_{\text{F}}$. Note
that the band gap $\Delta_0$ here is not the same with the observed transport band gap,
which also contains the contribution from electron correlations. The kinetic momentum of
a Dirac particle in an out-of-plane magnetic field is given by
\begin{eqnarray}
  \bsym{\pi}
  =
  \mathbf{p}-\frac{qB}{2}\mathbf{e}_{\perp}\times\mathbf{r},
\end{eqnarray}
where $\mathbf{p}=-\mathtt{i}\bsym{\nabla}$ is the momentum, $\mathbf{e}_{\perp}$ is a
unit vector perpendicular to the surface of the 2D system, and $B$ is the amplitude of the
out-of-plane magnetic field.

\subsection{Electromagnetic-field interaction\label{sec:electromagnetic}}

The electromagnetic-field interaction with the Dirac fermion can be included by extending
the kinetic momentum
\begin{eqnarray}
  \bsym{\pi}^{\star}(t)
  =
  \bsym{\pi}
  -q\int\tilde{\bsym{\mathcal{A}}}(\mathbf{k},t)e^{\mathtt{i}\mathbf{k}\cdot\mathbf{r}}
  \frac{\text{d}^2k}{(2\pi)^2},
\end{eqnarray}
where $\tilde{\bsym{\mathcal{A}}}(\mathbf{k},t)e^{\mathtt{i}\mathbf{k}\cdot\mathbf{r}}$ is
the vector potential in an electromagnetic mode with wavevector $\mathbf{k}$.  In the long-wavelength limit, the vector
potential can be simplified as
\begin{eqnarray}
  \tilde{\bsym{\mathcal{A}}}(\mathbf{k},t)e^{\mathtt{i}\mathbf{k}\cdot\mathbf{r}}
  \simeq
  \tilde{\bsym{\mathcal{A}}}(\mathbf{k},t)
  \left(1+\mathtt{i}\mathbf{k}\cdot\mathbf{r}+\cdots\right)
  \simeq
  \tilde{\bsym{\mathcal{A}}}(\mathbf{k},t).
\end{eqnarray}
Including the electromagnetic-field interaction, the electronic Hamiltonian can be rewritten as
\begin{eqnarray}
  \hat{\mathcal{H}}^{\star}(t)
  =
  \hat{\mathcal{H}}-\hat{\bsym{j}}\cdot\bsym{\mathcal{A}}(t),
\end{eqnarray}
where $\bsym{\mathcal{A}}(t)=\int\tilde{\bsym{\mathcal{A}}}(\mathbf{k},t)
\text{d}^2k/(2\pi)^2$ is the electromagnetic vector potential and
\begin{eqnarray}
  \hat{\bsym{j}}
  =
  q\sum_{\tau}\int\hat{\psi}^{\dagger}_{\tau}(\mathbf{r})\bsym{u}_{\tau}
  \hat{\psi}_{\tau}(\mathbf{r})\text{d}^2r
\end{eqnarray}
is the current operator with $\bsym{u}_{\tau}=v_{\text{F}}\bsym{\alpha}_{\tau}$ the
velocity matrix. We define $\text{d}\hat{\bsym{\xi}}(t)=\int_0^t  \hat{\bsym{j}}(t') dt'$ as the polarization operator.
It can be shown that the polarization operator can be related to the
current operator by
\begin{eqnarray}
  \frac{\text{d}\hat{\bsym{\xi}}}{\text{d}t}
  =
  \mathtt{i}\left[\hat{\mathcal{H}},\;\hat{\bsym{\xi}}\right]
  =
  \hat{\bsym{j}},
  \label{polarization_current}
\end{eqnarray}
where the interaction picture is used for the time-dependent fermion field operators. By
using a length-gauge transformation, the electromagnetic-field interaction in the
electronic Hamiltonian can be rewritten as a dipole-field interaction, and we have
\begin{eqnarray}
  \hat{\mathcal{H}}^{\star}_{\text{L}}(t)
  &=&
  e^{-\mathtt{i}\hat{\bsym{\xi}}\cdot\bsym{\mathcal{A}}(t)}
  \left[\hat{\mathcal{H}}^{\star}(t)-\mathtt{i}\frac{\partial}{\partial{t}}\right]
  e^{\mathtt{i}\hat{\bsym{\xi}}\cdot\bsym{\mathcal{A}}(t)}
  +\mathtt{i}\frac{\partial}{\partial{t}}\n
  &=&
  \hat{\mathcal{H}}
  -\hat{\bsym{\xi}}\cdot\bsym{\mathcal{F}}(t),
  \label{dipole_approximation}
\end{eqnarray}
with $\bsym{\mathcal{F}}=-{\partial{\bsym{\mathcal{A}}}}/{\partial{t}}$ being the
electromagnetic field. The formula of electromagnetic-field interaction in
Eq.~(\ref{dipole_approximation}) is known as the electric-dipole approximation. On the
other hand, if the polarization matrix element is difficult to solve, the current matrix
element can also be used to calculate the transition probabilities. The current matrix
element can be related to the polarization matrix element by using
Eq.~(\ref{polarization_current}), as
\begin{eqnarray}
  \langle N|\hat{\bsym{j}}|M\rangle
  &=&
  \langle N|\frac{\text{d}\hat{\bsym{\xi}}}{\text{d}t}|M\rangle
  =
  \mathtt{i}(E_N-E_M)\langle N|\hat{\bsym{\xi}}|M\rangle,
\end{eqnarray}
with $|N \langle$ the $N$-th state wavefunction and ${E}_{N}$ the $N$-th state eigenenergy.
Therefore, the transition probabilities can be related to the current matrix element via the relation
${\xi}^{\mu}_{NM}=-\mathtt{i}{j}^{\mu}_{NM}/({E_{N}-E_{M}})$, where
$\xi^{\mu}_{NM}=\langle{N}|\hat{\xi}^{\mu}|{M}\rangle$ is the polarization matrix element and
$j^{\mu}_{NM}=\langle{N}|\hat{j}^{\mu}|{M}\rangle$ is the current matrix element. It is what we
will do to solve one-exciton transition probability in Sec.~\ref{sec:valley_exciton},
because it is more convenient to find the current matrix element by the FW-transformed
formulations.

Based on Fermi's golden rule, the one-photon transition probability is given
by\cite{berkelbach2015bright}
\begin{eqnarray}
  \Gamma^{\mu}_{\text{1P}}(\omega)
  &=&
  {2\pi}\sum_{NM}\varrho_{M}|\xi^{\mu}_{NM}|^2\delta(\omega-E_{N}+E_{M}),
  \label{one_photon}
\end{eqnarray}
where $\varrho_M = \exp\left({-\beta_{\vartheta}{E}_{M}}\right)/
\left[\sum_{M'}\exp\left({-\beta_{\vartheta}\mathcal{E}_{M'}}\right)\right]$ is the
distribution function of the initial state with $\beta_{\vartheta}$ the inverse
temperature. The two-photon transition probability is given by the Kramers-Heisenberg
formula\cite{berkelbach2015bright}
\begin{eqnarray}
  \Gamma^{\mu}_{\text{2P}}(2\omega)
  &=&
  2\pi\sum_{NM}\varrho_{M}|\Upsilon^{\mu}_{NM}(\omega)|^2
  \delta(2\omega-E_{N}+E_{M}),\n
  \label{two_photon}
\end{eqnarray}
where $\Upsilon^{\mu}_{NM}(\omega) =
\sum_{Q}{\xi^{\mu}_{NQ}\xi^{\mu}_{QM}}/({E_{Q}-E_{M}-\omega-\mathtt{i}\eta})$ is the
two-photon transition amplitude with $\eta$ a line-broadening factor. By using these
formulations, the one-photon and two-photon absorption spectra can be studied.

\subsection{Electron-hole representation\label{sec:e_h_representation}}

The Fourier transformed single-particle Hamiltonian and fermion field operator are defined
as $\tilde{\mathtt{h}}_{\tau}(\mathbf{k})=\int\mathtt{h}_{\tau}(\mathbf{r})
e^{-\mathtt{i}\mathbf{k}\cdot\mathbf{r}}\text{d}^2r$,
$\hat{\tilde{\psi}}_{\tau}(\mathbf{k})=\int\hat{\psi}_{\tau}(\mathbf{r})
e^{-\mathtt{i}\mathbf{k}\cdot\mathbf{r}}\text{d}^2r$. It can be found that there is an
unitary transformation matrix such that
\begin{eqnarray}
  {\mathcal{U}}_{\tau}(\mathbf{k})\tilde{\mathtt{h}}_{\tau}(\mathbf{k})
  {\mathcal{U}}^{\dagger}_{\tau}(\mathbf{k})
  =
  \tilde{\varepsilon}_{\tau}(\mathbf{k})\beta,
  \label{diagonalization}
\end{eqnarray}
\begin{eqnarray}
  \hat{\tilde{\psi}}_{\tau}(\mathbf{k})
  &=&
  {\mathcal{U}}^{\dagger}_{\tau}(\mathbf{k})
  \begin{pmatrix}
    \hat{c}_{\tau}(\mathbf{k})\\ \hat{d}^{\dagger}_{\tau}(-\mathbf{k})
  \end{pmatrix},
\end{eqnarray}
\begin{eqnarray}
  \hat{\tilde{\psi}}^{\dagger}_{\tau}(\mathbf{k})
  &=&
  \begin{pmatrix}
    \hat{c}^{\dagger}_{\tau}(\mathbf{k}) & \hat{d}_{\tau}(-\mathbf{k})
  \end{pmatrix}
  {\mathcal{U}}_{\tau}(\mathbf{k}),
\end{eqnarray}
where $\tilde{\varepsilon}_{\tau}(\mathbf{k})$ is the single-particle energy,
$\hat{c}^{\dagger}_{\tau}(\mathbf{k})$, $\hat{c}_{\tau}(\mathbf{k})$ are the electron
creation, annihilation operators and $\hat{d}^{\dagger}_{\tau}(\mathbf{k})$,
$\hat{d}_{\tau}(\mathbf{k})$ are the hole creation, annihilation operators. The kinetic
part of the many-body Hamiltonian can be transformed as
\begin{eqnarray}
  \hat{\mathcal{H}}_{\text{kinetic}}
  &=&
  \sum_{\tau}\int\hat{\psi}^{\dagger}_{\tau}(\mathbf{r})
  \mathtt{h}_{\tau}(\mathbf{r})\hat{\psi}_{\tau}(\mathbf{r})\text{d}^2r\n
  &=&
  E_0+\sum_{\tau}\int\tilde{\varepsilon}_{\tau}(\mathbf{k})
  \big[\hat{c}^{\dagger}_{\tau}(\mathbf{k})\hat{c}_{\tau}(\mathbf{k})\n
  &&+\hat{d}^{\dagger}_{\tau}(-\mathbf{k})\hat{d}_{\tau}(-\mathbf{k})\big]
  \frac{\text{d}^2k}{(2\pi)^2},
\end{eqnarray}
where $E_0$ is the ground-state energy.

The polarization and current operators can be rewritten by integrations in quasi-momentum
space as
\begin{eqnarray}
  \hat{\bsym{\xi}}
  =
  \mathtt{i}q\sum_{\tau}\int\hat{\tilde{\psi}}^{\dagger}_{\tau}(\mathbf{k})
  \bsym{\nabla}_{\mathbf{k}}
  \hat{\tilde{\psi}}_{\tau}(\mathbf{k})\frac{\text{d}^2k}{(2\pi)^2},
\end{eqnarray}
\begin{eqnarray}
  \hat{\bsym{j}}
  =
  q\sum_{\tau}\int\hat{\tilde{\psi}}^{\dagger}_{\tau}(\mathbf{k})\bsym{u}_{\tau}
  \hat{\tilde{\psi}}_{\tau}(\mathbf{k})\frac{\text{d}^2k}{(2\pi)^2}.
\end{eqnarray}
Because the hole can be seen as the electron with negative energy in the present
formulation, the electron with charge $q=-e$ is chosen for the polarization and current
operators. The momentum matrix elements are defined by
\begin{eqnarray}
  \begin{pmatrix}
    \tilde{\bsym{\mathcal{P}}}_{\text{ee},\tau}(\mathbf{k})
    & \tilde{\bsym{\mathcal{P}}}_{\text{eh},\tau}(\mathbf{k})\\
    \tilde{\bsym{\mathcal{P}}}_{\text{he},\tau}(\mathbf{k})
    & \tilde{\bsym{\mathcal{P}}}_{\text{hh},\tau}(\mathbf{k})
  \end{pmatrix}
  =
  {\mathcal{U}}_{\tau}(\mathbf{k})\bsym{u}_{\tau}
  {\mathcal{U}}^{\dagger}_{\tau}(\mathbf{k}),
  \label{transformation_of_velocity}
\end{eqnarray}
and then the current operator can be rewritten as
\begin{eqnarray}
  \hat{\bsym{j}}
  &=&
  -e\sum_{\tau}\int\Big[
  \hat{c}^{\dagger}_{\tau}(\mathbf{k})
  \tilde{\bsym{\mathcal{P}}}_{\text{ee},\tau}(\mathbf{k})
  \hat{c}_{\tau}(\mathbf{k})\n
  &&
  +
  \hat{d}_{\tau}(-\mathbf{k})
  \tilde{\bsym{\mathcal{P}}}_{\text{he},\tau}(\mathbf{k})
  \hat{c}_{\tau}(\mathbf{k})
  +
  \hat{c}^{\dagger}_{\tau}(\mathbf{k})
  \tilde{\bsym{\mathcal{P}}}_{\text{eh},\tau}(\mathbf{k})
  \hat{d}^{\dagger}_{\tau}(-\mathbf{k})\n
  &&
  -
  \hat{d}^{\dagger}_{\tau}(-\mathbf{k})
  \tilde{\bsym{\mathcal{P}}}_{\text{hh},\tau}(\mathbf{k})
  \hat{d}_{\tau}(-\mathbf{k})\Big]\frac{\text{d}^2k}{(2\pi)^2}.
\end{eqnarray}
The dipole-moment matrix elements are defined by
\begin{eqnarray}
  \begin{pmatrix}
    \tilde{\bsym{\mathcal{D}}}_{\text{ee},\tau}(\mathbf{k})
    & \tilde{\bsym{\mathcal{D}}}_{\text{eh},\tau}(\mathbf{k})\\
    \tilde{\bsym{\mathcal{D}}}_{\text{he},\tau}(\mathbf{k})
    & \tilde{\bsym{\mathcal{D}}}_{\text{hh},\tau}(\mathbf{k})
  \end{pmatrix}
  =
  \mathtt{i}\bsym{\nabla}_{\mathbf{k}}
  +{\mathcal{U}}_{\tau}(\mathbf{k})\mathtt{i}\bsym{\nabla}_{\mathbf{k}}
  {\mathcal{U}}^{\dagger}_{\tau}(\mathbf{k}),
\end{eqnarray}
and the polarization operator becomes
\begin{eqnarray}
  \hat{\bsym{\xi}}
  &=&
  -e\sum_{\tau}\int\Big[
  \hat{c}^{\dagger}_{\tau}(\mathbf{k})
  \tilde{\bsym{\mathcal{D}}}_{\text{ee},\tau}(\mathbf{k})
  \hat{c}_{\tau}(\mathbf{k})\n
  &&+
  \hat{d}_{\tau}(-\mathbf{k})
  \tilde{\bsym{\mathcal{D}}}_{\text{he},\tau}(\mathbf{k})
  \hat{c}_{\tau}(\mathbf{k})
  +
  \hat{c}^{\dagger}_{\tau}(\mathbf{k})
  \tilde{\bsym{\mathcal{D}}}_{\text{eh},\tau}(\mathbf{k})
  \hat{d}^{\dagger}_{\tau}(-\mathbf{k})\n
  &&-
  \hat{d}^{\dagger}_{\tau}(-\mathbf{k})
  \tilde{\bsym{\mathcal{D}}}_{\text{hh},\tau}(\mathbf{k})
  \hat{d}_{\tau}(-\mathbf{k})\Big]\frac{\text{d}^2k}{(2\pi)^2}.
\end{eqnarray}
By using the electron-hole representation, it is convenient to write the excitonic excited
states, and the optical transition amplitudes can also be formulated.

Based on the electron-hole representation, an one-exciton excited state, also known as an
exciton, can be written as
\begin{eqnarray}
  |{\text{X}_{I\tau}}\rangle
  =
  \int\tilde{\Psi}_{\text{X},I\tau}(\mathbf{k})
  \hat{c}^{\dagger}_{\tau}(\mathbf{k})
  \hat{d}^{\dagger}_{\tau}(-\mathbf{k})
  \frac{\text{d}^2k}{(2\pi)^2}|{0}\rangle,
\end{eqnarray}
where the exciton wavefunction ${\Psi}_{\text{X},I\tau}(\mathbf{k})$ can be solved from
the eigenvalue equation
\begin{eqnarray}
  \int
  \tilde{\mathcal{H}}_{\text{X},\tau}(\mathbf{k},\mathbf{k}')
  \tilde{\Psi}_{\text{X},I\tau}(\mathbf{k}')\frac{\text{d}^2k'}{(2\pi)^2}
  =
  E_{\text{X},I\tau}\tilde{\Psi}_{\text{X},I\tau}(\mathbf{k}),
\end{eqnarray}
and the exciton Hamiltonian can be derived from
\begin{eqnarray}
  \tilde{\mathcal{H}}_{\text{X},\tau}(\mathbf{k},\mathbf{k}')
  &=&
  |{0}\rangle{{0}}|\hat{d}_{\tau}(-\mathbf{k})\hat{c}_{\tau}(\mathbf{k})
  \hat{\mathcal{H}}\hat{c}^{\dagger}_{\tau}(\mathbf{k}')
  \hat{d}^{\dagger}_{\tau}(-\mathbf{k}')|{0}\rangle.
\end{eqnarray}
The transition between an one-exciton excited state and the ground state is known as the
``one-exciton transition'', and the transition amplitude can be calculated in terms of either $\xi^{\mu}_{N0}$ or $ j^{\mu}_{N0}$, where
\begin{eqnarray}
  \xi^{\mu}_{N0}
  =
  -e\int\tilde{\Psi}^{*}_{\text{X},I_N\tau_N}(\mathbf{k})
  \tilde{\mathcal{D}}^{\mu}_{\text{eh},\tau_{N}}(\mathbf{k})\frac{\text{d}^2k}{(2\pi)^2}
  \label{transition_amplitude0}
\end{eqnarray}
and
\begin{eqnarray}
  j^{\mu}_{N0}
  =
  -e\int\tilde{\Psi}^{*}_{\text{X},I_{N}\tau_{N}}(\mathbf{k})
  \tilde{\mathcal{P}}^{\mu}_{\text{eh},\tau_{N}}(\mathbf{k})
  \frac{\text{d}^2k}{(2\pi)^2}.
  \label{transition_amplitude1}
\end{eqnarray}
The transition between different one-exciton excited states is known as the
``intra-exciton transition'' and the transition amplitudes can be  calculated in terms of either $\xi^{\mu}_{NM}$ or $j^{\mu}_{NM}$, where
\begin{eqnarray}
  \xi^{\mu}_{NM}
  &=&
  -e\int\tilde{\Psi}^{*}_{\text{X},I_N\tau_N}(\mathbf{k})
  \tilde{\mathcal{D}}^{\mu}_{\text{X},\tau_N}(\mathbf{k})
  \tilde{\Psi}_{\text{X},I_M\tau_N}(\mathbf{k})\frac{\text{d}^2k}{(2\pi)^2}\n
  \label{intra_exciton_transition_amplitude}
\end{eqnarray}
and
\begin{eqnarray}
  j^{\mu}_{NM}
  &=&
  -e\int\tilde{\Psi}^{*}_{\text{X},I_N\tau_N}(\mathbf{k})
  \tilde{\mathcal{P}}^{\mu}_{\text{X},\tau_N}(\mathbf{k})
  \tilde{\Psi}_{\text{X},I_M\tau_N}(\mathbf{k})\frac{\text{d}^2k}{(2\pi)^2}\n
\end{eqnarray}
with $\tilde{\mathcal{D}}^{\mu}_{\text{X},\tau}(\mathbf{k}) =
\tilde{\mathcal{D}}^{\mu}_{\text{ee},\tau}(\mathbf{k})
-\tilde{\mathcal{D}}^{\mu}_{\text{hh},\tau}(\mathbf{k})$ and
$\tilde{\mathcal{P}}^{\mu}_{\text{X},\tau}(\mathbf{k}) =
\tilde{\mathcal{P}}^{\mu}_{\text{ee},\tau}(\mathbf{k})
-\tilde{\mathcal{P}}^{\mu}_{\text{hh},\tau}(\mathbf{k})$. The ground-state energy can be
assigned as $E_{0}=0$ and the excited-state energy for a one-exciton excited state is
given by $E_{N}=E_{\text{X},I_N\tau_N}$. With the relations given above optical spectra
involving one-exciton transitions and intra-exciton transitions can be
calculated.

\section{FW transformation\label{sec:fw_transformation}}

Based on the derivation in Sec.~\ref{sec:e_h_representation}, we find that the key to connect
the Dirac equation and the exciton Hamiltonian is the unitary transformation matrix
${\mathcal{U}}(\mathbf{k})$. We need to solve the diagonalization problem in
Eq.~(\ref{diagonalization}) first to find the electron and hole energies. However, it
will be difficult if external fields are included in the Dirac Hamiltonian matrix. One
way to solve this problem is to replace the wavevector $\mathbf{k}$ by
$-\mathtt{i}\bsym{\nabla}$ in Eq.~(\ref{diagonalization}). Namely,
$\tilde{\mathtt{h}}_{\tau}(-\mathtt{i}\bsym{\nabla})\equiv \mathtt{h}_{\tau}(\mathbf{r})$ and
$\tilde{\varepsilon}_{\tau}(-\mathtt{i}\bsym{\nabla})\equiv{\varepsilon}_{\tau}(\mathbf{r})$.
We get
\begin{eqnarray}
  {\mathcal{U}}_{\tau}(-\mathtt{i}\bsym{\nabla})
  \mathtt{h}_{\tau}(\mathbf{r})
  {\mathcal{U}}^{\dagger}_{\tau}(-\mathtt{i}\bsym{\nabla})
  =
  {\varepsilon}_{\tau}(\mathbf{r})\beta.
  \label{unitary_transformation}
\end{eqnarray}
Eq.~(\ref{unitary_transformation}) is the basic idea of FW transformation. By a proper
choice of the unitary transformation, the analytic formulation of the single-particle
Hamiltonian $\varepsilon_{\tau}(\mathbf{r})$ can be derived. Since
$\varepsilon_{\tau}(\mathbf{r})\beta$ is still a functional of $-\mathtt{i}\bsym{\nabla}$,
 $\varepsilon_{\tau}(\mathbf{r})$ is actually the electron
Hamiltonian and $-\varepsilon_{\tau}(\mathbf{r})$ is the hole Hamiltonian.

\subsection{Eriksen method}

The FW transformation connecting the initial Hamiltonian $\mathcal{H}$ and the transformed
Hamiltonian $\mathcal{H}_{\text{FW}}$ can be written as\cite{foldy1950dirac,
eriksen1958foldy}
\begin{eqnarray}
  \mathcal{H}_{\text{FW}}
  =
  \mathcal{U}_{\text{FW}}\left(\mathcal{H}
  -\mathtt{i}\frac{\partial}{\partial{t}}\right)\mathcal{U}^{\dagger}_{\text{FW}}
  +\mathtt{i}\frac{\partial}{\partial{t}}.
\end{eqnarray}
where $\mathcal{U}_{\text{FW}}$ is an unitary operator named as the FW transformation
operator. Based on Eriksen's method\cite{eriksen1958foldy, silenko2008foldy,
silenko2016general}, we assume that the initial Hamiltonian can be divided into
\begin{eqnarray}
  \mathcal{H}={\beta}mv^2_{\text{F}}+\mathcal{E}+\mathcal{O}+\mathcal{O}',
\end{eqnarray}
with ${\beta}\mathcal{E}=\mathcal{E}{\beta}$, ${\beta}\mathcal{O}=-\mathcal{O}{\beta}$,
${\beta}\mathcal{O}'=-\mathcal{O}'{\beta}$, where $\mathcal{E}$ is an even operator and
$\mathcal{O}$ is an odd operator in the Hamiltonian, $\mathcal{O}'$ is an odd operator of
the external interaction of the Hamiltonian. Accordingly, the FW transformation operator
up to the order of $1/m^2$ has the expression\cite{silenko2008foldy, silenko2016general}
\begin{eqnarray}
  \mathcal{U}_{\text{FW}}
  &=&
  \exp\left(\frac{1}{2}\left[\mathcal{S},\;\mathcal{S}'\right]\right)
  \exp\left(\mathtt{i}\mathcal{S}'\right)
  \exp\left(\mathtt{i}\mathcal{S}\right),
\end{eqnarray}
where $\mathcal{S}=-\mathtt{i}\beta\mathcal{O}/(2mv^2_{\text{F}})$,  $\mathcal{S}' =
-\mathtt{i}\left[\mathcal{O},\mathcal{G}\right]/(4m^2v^4_{\text{F}})$,
with
\begin{eqnarray}
  \mathcal{G}=\mathcal{E}+\mathcal{O}'-\mathtt{i}\frac{\partial}{\partial{t}}.
  \label{generator}
\end{eqnarray}
The FW Hamiltonian up to the order of $1/m^2$ can be written as
\begin{eqnarray}
  \mathcal{H}_{\text{FW}}
  &=&
  {\beta}mv^2_{\text{F}}+\mathcal{E}+\mathcal{O}'
  +\frac{{\beta}\mathcal{O}^2}{2mv^2_{\text{F}}}
  -\frac{\left[\mathcal{O},\left[\mathcal{O},\mathcal{G}\right]\right]}
  {8m^2v^4_{\text{F}}}.
\end{eqnarray}
Note that the odd operator $\mathcal{O}'$ is not discussed by cited
literatures\cite{eriksen1958foldy, silenko2008foldy, silenko2016general}. As we will show
lately in Sec.~\ref{sec:interband_transition}, the operator contributes mainly to the
interband transition, which is not considered in atomic systems.

\subsection{Single-particle Hamiltonian}

In the present case, the correspondences between the single-particle Hamiltonian and the
Eriksen's Hamiltonian are given by
\begin{eqnarray}
  \mathcal{H}={\mathtt{h}}-q\mathbf{r}\cdot\bsym{\mathcal{F}},\hskip2ex
  \mathcal{E}=-q\mathbf{r}\cdot\left(\mathbf{F}+\bsym{\mathcal{F}}\right),
\end{eqnarray}
\begin{eqnarray}
  \mathcal{O}=v_{\text{F}}\bsym{\pi}\cdot{\bsym{\alpha}_{\tau}},\hskip2ex
  \mathcal{O}'=0.
\end{eqnarray}
By using $\bsym{\pi}\times\bsym{\pi} = \mathtt{i}qB\mathbf{e}_{\perp}$ and
$\left(\mathbf{a}\cdot\bsym{\alpha}_{\tau}\right)
\left(\mathbf{b}\cdot\bsym{\alpha}_{\tau}\right) = \mathbf{a}\cdot\mathbf{b}
+\mathtt{i}\tau\mathbf{e}_{\perp}\cdot\mathbf{a}\times\mathbf{b}\beta$, we find
$\mathcal{O}^2 = {v^2_{\text{F}}}\left(|\bsym{\pi}|^2-\beta\tau qB\right)$,
$\left[\mathcal{O},\mathcal{G}\right] = \mathtt{i}qv_{\text{F}}\left(\mathbf{F}
+\bsym{\mathcal{F}}\right)\cdot{\bsym{\alpha}_{\tau}}$, and
$\left[\mathcal{O},\left[\mathcal{O},\mathcal{G}\right]\right] = 2\tau
qv^2_{\text{F}}\beta\mathbf{e}_{\perp}
\cdot\left(\mathbf{F}+\bsym{\mathcal{F}}\right)\times\bsym{\pi}$. The FW Hamiltonian is
given by
\begin{eqnarray}
  \mathcal{H}_{\text{FW}}
  &=&
  \beta\left[mv^2_{\text{F}}+\frac{|\bsym{\pi}|^2}{2m}
  -\tau\frac{q\mathbf{e}_{\perp}\cdot\left(\mathbf{F}+\bsym{\mathcal{F}}\right)
  \times\bsym{\pi}}{4m^2v^2_{\text{F}}}\right]\n
  &&-q\left(\mathbf{F}+\bsym{\mathcal{F}}\right)\cdot\mathbf{r}-\tau\frac{qB}{2m}.
\end{eqnarray}
Based on the FW single-particle Hamiltonian, the effective single-particle energy is given
by
\begin{eqnarray}
  \varepsilon_{\text{FW},i}
  &=&
  mv^2_{\text{F}}+\frac{|\bsym{\pi}_{i}|^2}{2m}
  -\tau_{i}\frac{q_i\mathbf{e}_{\perp}
  \cdot\left(\mathbf{F}+\bsym{\mathcal{F}}\right)\times\bsym{\pi}_{i}}
  {4m^2v^2_{\text{F}}}\n
  &&-q_{i}\left(\mathbf{F}+\bsym{\mathcal{F}}\right)\cdot\mathbf{r}_{i}
  -\tau_{i}\frac{q_{i}B}{2m},
\end{eqnarray}
where $i\in\{\text{e},\text{h}\}$, $q=-e$ for electrons and $q=e$ for holes,
$\tau_{\text{e}}$ and $\tau_{\text{h}}$ are valley indices of the effective electron and
hole.

\subsection{Two-particle interaction}

Now we consider a two-particle Hamiltonian
\begin{eqnarray}
  \mathcal{H}
  =
  \mathtt{h}_1+\mathtt{h}_2+\mathcal{V}_{12},
\end{eqnarray}
where $\mathcal{V}_{12}$ is a two-particle interaction, and ${\mathtt{h}}_i =
v_{\text{F}}\bsym{\pi}_i\cdot{\bsym{\alpha}_{\tau}}_i+mv^2_{\text{F}}\beta_i+q_{i}\Phi_i$
is the single-particle Hamiltonian matrix. The Hamiltonian can be divided as
\begin{eqnarray}
  \mathcal{H}
  =
  \beta_1mv^2_{\text{F}}+\beta_2mv^2_{\text{F}}+\mathcal{E}
  +\mathcal{O}_1+\mathcal{O}_2,
\end{eqnarray}
\begin{eqnarray}
  \mathcal{E}=q_1\Phi_1+q_2\Phi_2+\mathcal{V}_{12},
\end{eqnarray}
\begin{eqnarray}
  \mathcal{O}_1=v_{\text{F}}\bsym{\pi}_1\cdot{\bsym{\alpha}_{\tau_1}},\hskip2ex
  \mathcal{O}_2=v_{\text{F}}\bsym{\pi}_2\cdot{\bsym{\alpha}_{\tau_2}}.
\end{eqnarray}
The FW transformed Hamiltonian is then given by
\begin{eqnarray}
  \mathcal{H}_{\text{FW}}
  &=&
  \mathtt{h}_1+\mathtt{h}_2
  -\frac{\left[\mathcal{O}_1,\left[\mathcal{O}_1,\mathcal{G}\right]\right]}
  {8m^2v^4_{\text{F}}}
  -\frac{\left[\mathcal{O}_2,\left[\mathcal{O}_2,\mathcal{G}\right]\right]}
  {8m^2v^4_{\text{F}}}\n
  &&-\frac{\left[\mathcal{O}_1,\left[\mathcal{O}_2,\mathcal{G}\right]\right]
  +\left[\mathcal{O}_2,\left[\mathcal{O}_1,\mathcal{G}\right]\right]}
  {8m^2v^4_{\text{F}}},
\end{eqnarray}
with $\mathcal{G}=\mathcal{V}_{12}-\mathtt{i}\partial/\partial{t}$. It can shown $\left[\mathcal{O}_{i},\mathcal{G}\right] =
-\mathtt{i}v_{\text{F}}\bsym{\alpha}_{\tau_i}\cdot\bsym{\nabla}_{i}\mathcal{V}_{12}$,
$\left[\mathcal{O}_i,\left[\mathcal{O}_i,\mathcal{G}\right]\right] =
-v^2_{\text{F}}\nabla^2_i\mathcal{V}_{12} -2v^2_{\text{F}}\beta_i\tau_i
\mathbf{e}_{\perp}\cdot\bsym{\nabla}_i\mathcal{V}_{12}\times\bsym{\pi}_i$, and
$\left[\mathcal{O}_{i},\left[\mathcal{O}_{j},\mathcal{G}\right]\right] = -v^2_{\text{F}}
(\bsym{\alpha}_{\tau_i}\cdot\bsym{\nabla}_{i})
(\bsym{\alpha}_{\tau_j}\cdot\bsym{\nabla}_{j})\mathcal{V}_{12}$ for $i\neq{j}$. The FW
transformed two-particle interaction is given by
\begin{eqnarray}
  \mathcal{V}_{\text{FW},12}
  &=&
  \mathcal{V}_{12}
  +\beta_1\tau_1
  \frac{\mathbf{e}_{\perp}\cdot\bsym{\nabla}_1\mathcal{V}_{12}\times\bsym{\pi}_1}
  {4m^2v^2_{\text{F}}}+\frac{\nabla^2_1\mathcal{V}_{12}}{8m^2v^2_{\text{F}}}\n
  &&+\beta_2\tau_2
  \frac{\mathbf{e}_{\perp}\cdot\bsym{\nabla}_2\mathcal{V}_{12}\times\bsym{\pi}_2}
  {4m^2v^2_{\text{F}}}+\frac{\nabla^2_2\mathcal{V}_{12}}{8m^2v^2_{\text{F}}}\n
  &&+\frac{\left(\bsym{\alpha}_{\tau_1}\cdot\bsym{\nabla}_{1}\right)
  \left(\bsym{\alpha}_{\tau_2}\cdot\bsym{\nabla}_{2}\right)\mathcal{V}_{12}}
  {4m^2v^2_{\text{F}}}.
\end{eqnarray}
For 2D materials, the two-particle interaction is assumed to be given by the screened
Coulomb interaction
\begin{eqnarray}
  \mathcal{V}_{ij}=\frac{q_iq_j}{e^2}V({r}_{ij}).
\end{eqnarray}
The FW transformed screened Coulomb potential is then given by
\begin{eqnarray}
  V_{\text{FW},ij}
  &=&
  V({r}_{ij})+\tau_{i}\frac{\mathbf{e}_{\perp}\cdot\bsym{\nabla}_{i}
  V({r}_{ij})\times\bsym{\pi}_{i}}{4m^2v^2_{\text{F}}}
  +\frac{\nabla^2_{i}V({r}_{ij})}{8m^2v^2_{\text{F}}}\n
  &&+\tau_{j}\frac{\mathbf{e}_{\perp}\cdot\bsym{\nabla}_{j}
  V({r}_{ij})\times\bsym{\pi}_{j}}{4m^2v^2_{\text{F}}}
  +\frac{\nabla^2_{j}V({r}_{ij})}{8m^2v^2_{\text{F}}}
\end{eqnarray}
for $i,j\in\{\text{e},\text{h}\}$. The FW transformed screened Coulomb potential can be
used to describe the electron-electron interaction, the electron-hole interaction, and the
hole-hole interaction in 2D materials.

\subsection{Interband transition\label{sec:interband_transition}}

Assuming that the contributions from the static electric and magnetic fields to the
interband transition can be neglected, the correspondences between electromagnetic-field
interaction and Eriksen's Hamiltonian are given by
\begin{eqnarray}
  \mathcal{H}=\mathtt{h}^{(0)}-qv_{\text{F}}\bsym{\alpha}_{\tau}\cdot\bsym{\mathcal{A}},
  \hskip2ex
  \mathcal{E}=0,\hskip2ex
  \mathcal{O}=v_{\text{F}}\mathbf{p}\cdot{\bsym{\alpha}_{\tau}},
\end{eqnarray}
\begin{eqnarray}
  \mathcal{O}'=-qv_{\text{F}}\bsym{\alpha}_{\tau}\cdot\bsym{\mathcal{A}}.
\end{eqnarray}
We find $\left[\mathcal{O},\mathcal{G}\right] =
2\mathtt{i}qv^2_{\text{F}}\tau\mathbf{e}_{\perp}\cdot\bsym{\mathcal{A}}
\times\mathbf{p}\beta$, $\left[\mathcal{O},\left[\mathcal{O},\mathcal{G}\right]\right] =
-4qv^3_{\text{F}}\left(\mathbf{e}_{\perp}\cdot\bsym{\mathcal{A}}\times\mathbf{p}\right)
(\mathbf{e}_{\perp}\cdot\bsym{\alpha}_{\tau}\times\mathbf{p})$. The FW transformed Hamitonian with
electromagnetic-field interaction is given by
\begin{eqnarray}
  \mathcal{H}_{\text{FW}}
  &=&
  \mathtt{h}^{(0)}_{\text{FW}}-q\bsym{u}_{\text{FW}}\cdot\bsym{\mathcal{A}},
\end{eqnarray}
where the FW effective velocity matrix is
\begin{eqnarray}
  \bsym{u}_{\text{FW}}\cdot\bsym{\mathcal{A}}
  &=&
  v_{\text{F}}\bsym{\alpha}_{\tau}\cdot\bsym{\mathcal{A}}
  -\frac{v_{\text{F}}\left(\mathbf{e}_{\perp}\cdot\bsym{\mathcal{A}}\times\mathbf{p}\right)
  (\mathbf{e}_{\perp}\cdot\bsym{\alpha}_{\tau}\times\mathbf{p})}{2m^2v^2_{\text{F}}}.\n
\end{eqnarray}
By defining the circular-polarized components of variables
$\mathcal{A}^{\pm}=(\mathcal{A}^{x}\pm\mathtt{i}\mathcal{A}^{y})/\sqrt{2}$,
$p^{\pm}=(p^{x}\pm\mathtt{i}p^{y})/\sqrt{2}$, and
\begin{eqnarray}
  \alpha^{\pm}
  &=&
  \frac{1}{\sqrt{2}}\left(\alpha^{x}\pm\mathtt{i}\alpha^{y}\right)
  =
  \frac{1}{\sqrt{2}}
  \begin{pmatrix}
    0 & {\tau\pm{1}}\\ {\tau\mp{1}} & 0
  \end{pmatrix},
\end{eqnarray}
we can find $\bsym{\alpha}_{\tau}\cdot\bsym{\mathcal{A}}=
\alpha^{+}\mathcal{A}^{-}+\alpha^{-}\mathcal{A}^{+}$ and
\begin{eqnarray}
  \left(\mathbf{e}_{\perp}\cdot\bsym{\mathcal{A}}\times\mathbf{p}\right)
  (\mathbf{e}_{\perp}\cdot\bsym{\alpha}_{\tau}\times\mathbf{p})
  &=&
  \left[\alpha^{-}p^2/2-\alpha^{+}(p^{-})^2\right]\mathcal{A}^{+}\n
  &&+\left[\alpha^{+}p^2/2-\alpha^{-}(p^{+})^2\right]\mathcal{A}^{-}.\n
\end{eqnarray}
The FW velocity matrix, ${u}^{\pm}_{\text{FW}} =({u}^{x}_{\text{FW}}
\pm\mathtt{i}{u}^{y}_{\text{FW}})/\sqrt{2}$, is given by
\begin{eqnarray}
  {u}^{\pm}_{\text{FW}}
  =
  \left(1-\frac{p^2}{4m^2v^2_{\text{F}}}\right)v_{\text{F}}\alpha^{\pm}
  +\frac{(p^{\pm})^2}{2m^2v^2_{\text{F}}}v_{\text{F}}\alpha^{\mp}.
  \label{velocity_matrix}
\end{eqnarray}
The velocity matrix will be used to derive the selection rule and transition amplitude in
Sec.~\ref{sec:exciton_transitions}.

\section{Applications to valley excitons\label{sec:valley_exciton}}

In this section, we will use the formulation derived from the FW transformation of a
gapped Dirac Hamiltonian to study valley excitons in 2D materials. Based on the
formulations, the Hamiltonian for a valley exciton in external fields can be written. The
valley-exciton Hamiltonian is similar to the Wannier-exciton Hamiltonian, which is written
as an electron and a hole bound by an attractive Coulomb interaction, except that the
band-geometry correction terms are added. The effects of external-field interactions,
including electric-field interaction, electromagnetic-field interaction, and
magnetic-field interaction, on valley excitons can then be studied.

\subsection{Valley excitons in electric fields}

An exciton in an electric field ($|\mathbf{F}|\neq{0}$) and an electromagnetic field
($|\bsym{\mathcal{A}}|\neq{0}$) without magnetic field (${B}={0}$) is considered in this
section. The effective exciton Hamiltonian derived from the FW transformed Dirac
Hamiltonian is given by
\begin{eqnarray}
  \mathcal{H}_{\text{X}}
  &=&
  \delta\Delta
  +\left(\varepsilon_{\text{FW},\text{e}}+\delta\varepsilon_{\text{e}}\right)
  +\left(\varepsilon_{\text{FW},\text{h}}+\delta\varepsilon_{\text{h}}\right)
  -V_{\text{FW},\text{eh}}\n
  &=&
  \Delta+\frac{|\mathbf{p}_{\text{e}}|^2}{2m_{\text{e}}}
  +\frac{|\mathbf{p}_{\text{h}}|^2}{2m_{\text{h}}}-V({r}_{\text{eh}})\n
  &&+\tau_{\text{e}}\frac{\mathbf{e}_{\perp}\cdot\left[e
  \left(\mathbf{F}+\bsym{\mathcal{F}}\right)
  -\bsym{\nabla}_{\text{e}}V({r}_{\text{eh}})\right]
  \times\mathbf{p}_{\text{e}}}{4{m}^2v^2_{\text{F}}}\n
  &&-\tau_{\text{h}}\frac{\mathbf{e}_{\perp}\cdot\left[e
  \left(\mathbf{F}+\bsym{\mathcal{F}}\right)
  +\bsym{\nabla}_{\text{h}}V({r}_{\text{eh}})\right]
  \times\mathbf{p}_{\text{h}}}{4{m}^2v^2_{\text{F}}}\n
  &&+e(\mathbf{r}_{\text{e}}-\mathbf{r}_{\text{h}})\cdot
  \left(\mathbf{F}+\bsym{\mathcal{F}}\right)
  -\frac{\nabla^2_{\text{e}}V({r}_{\text{eh}})}{8{m}^2v^2_{\text{F}}}
  -\frac{\nabla^2_{\text{h}}V({r}_{\text{eh}})}{8{m}^2v^2_{\text{F}}},\n
\end{eqnarray}
where $\delta\Delta$ is the electron-correlation modification to the band-gap energy,
$\Delta=\delta\Delta+2mv^2_{\text{F}}$ is the band-gap energy,
$\delta\varepsilon_{\text{e}}$ and $\delta\varepsilon_{\text{h}}$ are band-asymmetry
modifications to the electron and hole energies, $m_{\text{e}}$ and $m_{\text{h}}$ are
effective electron mass and effective hole mass, which are defined by
\begin{eqnarray}
  \frac{|\mathbf{p}|^2}{2m_{\text{e}}}
  =
  \frac{|\mathbf{p}|^2}{2m}+\delta\varepsilon_{\text{e}},\hskip2ex
  \frac{|\mathbf{p}|^2}{2m_{\text{h}}}
  =
  \frac{|\mathbf{p}|^2}{2m}+\delta\varepsilon_{\text{h}}.
\end{eqnarray}
Assuming $\delta\varepsilon_{\text{e}}=-\delta\varepsilon_{\text{h}}$, we can find
$m=2\mu$, with $\mu=m_{\text{e}}m_{\text{h}}/(m_{\text{e}}+m_{\text{h}})$ the reduced
mass. A coordinate transformation to the relative coordinate system is applied
\begin{eqnarray}
  \mathbf{R}
  &=&
  \frac{m_{\text{e}}\mathbf{r}_{\text{e}}
  +m_{\text{h}}\mathbf{r}_{\text{h}}}{m_{\text{X}}},\hskip2ex
  \mathbf{r}
  =
  \mathbf{r}_{\text{e}}-\mathbf{r}_{\text{h}},
  \label{relative_coordinate_position}
\end{eqnarray}
\begin{eqnarray}
  \mathbf{p}_{\text{e}}
  =
  \frac{m_{\text{e}}}{m_{\text{X}}}\mathbf{P}+\mathbf{p},\hskip1ex
  \mathbf{p}_{\text{h}}
  =
  \frac{m_{\text{h}}}{m_{\text{X}}}\mathbf{P}-\mathbf{p},
  \label{relative_coordinate_momentum}
\end{eqnarray}
where $m_{\text{X}}=m_{\text{e}}+m_{\text{h}}$ is the exciton mass, $\mathbf{R}$ and
$\mathbf{P}=-\mathtt{i}\bsym{\nabla}_{\mathbf{R}}$ are the center-of-mass position and
momentum, $\mathbf{r}$ and $\mathbf{p}=-\mathtt{i}\bsym{\nabla}$ are internal-coordinate
position and momentum. The exciton Hamiltonian becomes
\begin{eqnarray}
  \mathcal{H}_{\text{X}}
  =
  \mathcal{H}'_{\text{X}}+\mathcal{H}''_{\text{X}}+\mathcal{J}_{\text{X}},
\end{eqnarray}
where
\begin{eqnarray}
  \mathcal{H}'_{\text{X}}
  &=&
  \frac{P^2}{2m_{\text{X}}}
  +\frac{e\Omega(m_{\text{e}}\tau_{\text{e}}-m_{\text{h}}\tau_{\text{h}})}
  {4m_{\text{X}}}
  \mathbf{e}_{\perp}\cdot\left(\mathbf{F}+\bsym{\mathcal{F}}\right)\times\mathbf{P}
  \label{translational_Hamiltonian}
\end{eqnarray}
is the exciton translational Hamiltonian,
\begin{eqnarray}
  \mathcal{H}''_{\text{X}}
  &=&
  \Delta+\frac{p^2}{2\mu}-V({r})
  +e\mathbf{r}\cdot\left(\mathbf{F}+\bsym{\mathcal{F}}\right)
  -\frac{\Omega}{4}\nabla^2V({r})\n
  &&+\frac{\Omega(\tau_{\text{e}}+\tau_{\text{h}})}{4}
  \mathbf{e}_{\perp}\cdot\left[e\left(\mathbf{F}+\bsym{\mathcal{F}}\right)
  -\bsym{\nabla}V({r})\right]\times\mathbf{p},
  \label{internal_hamiltonian0}
\end{eqnarray}
is the exciton internal Hamiltonian with
\begin{eqnarray}
  \Omega
  =
  \frac{1}{m^2v^2_{\text{F}}}
  =
  \frac{1}{4\mu^2v^2_{\text{F}}}
\end{eqnarray}
the Berry curvature, and
\begin{eqnarray}
  \mathcal{J}_{\text{X}}
  &=&
  -\frac{e\Omega(m_{\text{e}}\tau_{\text{e}}-m_{\text{h}}\tau_{\text{h}})}
  {4m_{\text{X}}}
  \mathbf{e}_{\perp}\cdot\bsym{\nabla}V({r})\times\mathbf{P}
\end{eqnarray}
is the exciton translational-internal coupling. For the exciton internal Hamiltonian, the
first four terms describe a Wannier exciton in an electric field, the fifth term is the
Darwin interaction, and the last term is the exciton valley-orbit coupling, which
resembles 2D version of spin-orbit coupling for atomistic systems. Both the exciton
valley-orbit coupling and the Darwin interaction can be considered as the contributions
from the Berry-curvature effect to the exciton energy levels of intravalley excitons
($\tau_{\text{e}}=\tau_{\text{h}}$).

Based on the translational Hamiltonian in Eq.~(\ref{translational_Hamiltonian}), it is
also shown that the Berry-curvature effect contributes to the valley-dependent exciton
transport and the exciton translational-internal coupling in the electric field for
intervalley excitons ($\tau_{\text{e}}=-\tau_{\text{h}}$). By using the Heisenberg
equation of motion and omitting the translational-internal coupling, the exciton mobility
in an electric field can be obtained by
\begin{eqnarray}
  \Big\langle\frac{\text{d}\mathbf{R}}{\text{d}t}\Big\rangle
  &=&
  -\mathtt{i}\langle\left[\mathbf{R},\mathcal{H}'_{\text{X}}\right]\rangle\n
  &=&
  \frac{\langle\mathbf{P}\rangle}{m_{\text{X}}}
  +
  \frac{e\Omega(m_{\text{e}}\tau_{\text{e}}-m_{\text{h}}\tau_{\text{h}})}
  {4m_{\text{X}}}\mathbf{e}_{\perp}\times\mathbf{F},
  \label{exciton_mobility}
\end{eqnarray}
where $\langle\mathbf{P}\rangle=0$ for the ensemble of excitons. The second term in
Eq.~(\ref{exciton_mobility}) can be seen as a contribution to the exciton valley Hall
effect, which has been observed experimentally\cite{onga2017exciton}. This is effect is believed to be
driven by the side-jump together with the skew-scattering mechanism\cite{glazov2020skew}.
While the contribution in Eq.~(\ref{exciton_mobility}) to the exciton valley Hall effect
is infinitesimal for intravalley excitons, it may be measurable if intervalley excitons
also contribute to the exciton mobility. Due to the simplicity of the present
model, the contribution of intervalley excitons to the exciton Hall
conductivity will be studied by an improved model in future research.

\subsection{Exciton energy-level splitting}

\begin{table*}
\centering
\begin{tabular}{c| c c c c c c c | c c c c c c}
\hline
Materials & $\Delta$(eV) & $a$(\AA) & $t$(eV)
& $m_{\text{e}}/m_0$ & $m_{\text{h}}/m_0$ & $r_0$(\AA) & $\kappa$
& $1s$ & $2s$ & $3s$
& $2p_{+}/2p_{-}$ & $3p_{+}/3p_{-}$ & $3d_{+}/3d_{-}$\\
\hline\hline
\multirow{2}{*}{$\text{MoS}_2$}
      & \multirow{2}{*}{$1.918$}
      & \multirow{2}{*}{$3.193$}
      & \multirow{2}{*}{$1.0799$}
      & \multirow{2}{*}{$0.47$}
      & \multirow{2}{*}{$0.54$}
      & \multirow{2}{*}{$44.68$}
      & $1.0$
      & $506.2$
      & $243.3$
      & $151.3$
      & $292.4/303.0$
      & $174.6/178.8$
      & $197.7/202.5$ \\
      & & & &  &  &  & $2.0$
      & $333.6$
      & $124.4$
      & $66.1$
      & $156.7/163.4$
      & $78.6/80.8$
      & $89.8/92.0$ \\
\hline
\multirow{2}{*}{$\text{MoSe}_2$}
      & \multirow{2}{*}{$1.516$}
      & \multirow{2}{*}{$3.313$}
      & \multirow{2}{*}{$0.8984$}
      & \multirow{2}{*}{$0.55$}
      & \multirow{2}{*}{$0.59$}
      & \multirow{2}{*}{$53.16$}
      & $1.0$
      & $458.7$
      & $230.5$
      & $147.6$
      & $274.0/284.1$
      & $169.0/173.2$
      & $190.5/195.3$ \\
      & & & &  &  &  & $2.0$
      & $309.2$
      & $122.8$
      & $67.6$
      & $152.9/159.7$
      & $79.9/82.2$
      & $91.4/93.8$ \\
\hline
\multirow{2}{*}{$\text{WS}_2$}
      & \multirow{2}{*}{$2.042$}
      & \multirow{2}{*}{$3.197$}
      & \multirow{2}{*}{$1.3315$}
      & \multirow{2}{*}{$0.32$}
      & \multirow{2}{*}{$0.35$}
      & \multirow{2}{*}{$40.17$}
      & $1.0$
      & $488.2$
      & $215.8$
      & $127.0$
      & $263.8/274.2$
      & $148.3/152.3$
      & $169.0/173.4$ \\
      & & & &  &  &  & $2.0$
      & $308.2$
      & $102.4$
      & $51.1$
      & $130.8/136.6$
      & $61.1/62.9$
      & $69.4/71.1$\\
\hline
\multirow{3}{*}{$\text{WSe}_2$}
      & \multirow{3}{*}{$1.761$}
      & \multirow{3}{*}{$3.310$}
      & \multirow{3}{*}{$1.1395$}
      & \multirow{2}{*}{$0.34$}
      & \multirow{2}{*}{$0.36$}
      & \multirow{2}{*}{$47.57$}
      & $1.0$
      & $437.2$
      & $200.6$
      & $120.9$
      & $242.9/253.2$
      & $140.3/144.3$
      & $159.5/164.0$ \\
      & & & &  &  &  & $2.0$
      & $281.2$
      & $98.3$
      & $50.4$
      & $124.5/130.6$
      & $60.0/61.9$
      & $68.4/70.2$ \\
      & & & & $0.38$ & $0.43$ & $46.80$ & $4.0$
      & $172.0$
      & $43.7$
      & $19.4$
      & $57.9/60.1$
      & $23.5/24.0$
      & $25.5/25.9$ \\
 \hline
\end{tabular}
\caption{Parameters of TMDCs ($\Delta$ the band-gap energy; $a$ the lattice constant; $t$
the hopping coupling; $r_0$ the screening length; $\kappa$ the dielectric constant) and
calculated exciton binding energies (in meV) of intravalley excitons with $\tau=1$. The
Fermi velocity is given by $v_{\text{F}}=at$. The parameters are obtained from
Ref.~\onlinecite{wu2019exciton} except the last row. The parameters ($m_{\text{e}}/m_0$,
$m_{\text{h}}/m_0$, $r_0$(\AA), $\kappa$) in the last row are chosen to fit the
experimentally observed photoluminescence spectra reported in
Ref.~\onlinecite{liu2019magnetophotoluminescence}. The electron-hole exchange effect has been included in the last row according to Ref.~\cite{mypaper0}.}
\label{tab:TMDCs}
\end{table*}

In this section, exciton internal Hamiltonians without external fields are considered, and
the eigenspectrum are used to study the Berry-curvature effect on exciton energy levels.
By assuming that $\tau_{\text{e}}=\tau_{\text{h}}=\tau$, the exciton internal Hamiltonian
for the intravalley exciton can be written as
\begin{eqnarray}
  \mathcal{H}''_{\text{intravalley-X},\tau}(\mathbf{r})
  &=&
  \Delta+\frac{p^2}{2\mu}-V({r})-\frac{\Omega}{4}\nabla^2V({r})\n
  &&-\frac{\tau\Omega}{2}
  \mathbf{e}_{\perp}\cdot\bsym{\nabla}V({r})\times\mathbf{p}.
  \label{intra_exciton_internal_Hamiltonian}
\end{eqnarray}
On the other hand, by assuming that $\tau_{\text{e}}=-\tau_{\text{h}}=\tau$, the exciton
internal Hamiltonian for the intervalley exciton can be written as
\begin{eqnarray}
  \mathcal{H}''_{\text{intervalley-X},\tau}(\mathbf{r})
  &=&
  \Delta+\frac{p^2}{2\mu}-V({r})-\frac{\Omega}{4}\nabla^2V({r}).
  \label{inter_exciton_internal_Hamiltonian}
\end{eqnarray}
The Darwin interaction is found in both Hamiltonians, but only the intravalley-exciton
Hamiltonian contains the exciton valley-orbit coupling. The exciton eigenenergy and
wavefunction can be calculated by the eigenvalue equation
\begin{eqnarray}
  \mathcal{H}''_{\text{X},\tau}(\mathbf{r})\Psi_{\text{X},I\tau}(\mathbf{r})
  =
  E_{\text{X},I\tau}\Psi_{\text{X},I\tau}(\mathbf{r}),
\end{eqnarray}
for both intravalley excitons and intervalley excitons. Since the exciton internal
Hamiltonian and the angular momentum operator commutes,
$\left[\mathcal{H}''_{\text{X},\tau}(\mathbf{r}), \;\mathcal{L}(\mathbf{r})\right]=0$,
with the angular momentum operator being given by
\begin{eqnarray}
  \mathcal{L}(\mathbf{r})
  =
  -\mathtt{i}\mathbf{e}_{\perp}\cdot\mathbf{r}\times\bsym{\nabla}
  =
  -\mathtt{i}\frac{\partial}{\partial\varphi},
\end{eqnarray}
the exciton wavefunction is also an eigenfunction of the angular momentum operator,
$\mathcal{L}(\mathbf{r}) \Psi_{\text{X},I\tau}(\mathbf{r}) =
l\Psi_{\text{X},I\tau}(\mathbf{r})$, with $l$ the angular momentum of the exciton. The
exciton wavefunction can be written as $\Psi_{\text{X},I\tau}(\mathbf{r}) =
e^{\mathtt{i}l\varphi}\mathcal{R}_{n}(r)$, where $\mathcal{R}_{n}(r)$ is the radial
wavefunction of the exciton, with $n$ the principal quantum number. By using
$\mathbf{p}=-\mathtt{i}\bsym{\nabla}$ and
\begin{eqnarray}
  \mathbf{e}_{\perp}\cdot\bsym{\nabla}V(r)\times\bsym{\nabla}
  =
  \frac{1}{r}\frac{\partial V(r)}{\partial{r}}
  \mathbf{e}_{\perp}\cdot\mathbf{r}\times\bsym{\nabla},
\end{eqnarray}
it can be shown that the exciton vally-orbit coupling ($\mathcal{V}_{\text{X,VOC}}$), the
fifth term in Eq.~(\ref{intra_exciton_internal_Hamiltonian}) and
Eq.~(\ref{inter_exciton_internal_Hamiltonian}) is proportional to the angular momentum
operator
\begin{eqnarray}
  \mathcal{V}_{\text{X,VOC}}(\mathbf{r})
  \propto
  -\frac{\Omega}{2}\frac{1}{r}\frac{\partial V(r)}{\partial{r}}\mathcal{L}(\mathbf{r}),
\end{eqnarray}
such that $\langle \mathcal{V}_{\text{X,VOC}}\rangle_{I\tau}\propto{l}$ with
\begin{eqnarray}
  \langle\mathcal{O}\rangle_{I\tau}
  \equiv
  \int\Psi^*_{\text{X},I\tau}(\mathbf{r})
  \mathcal{O}_{\tau}(\mathbf{r})\Psi_{\text{X},I\tau}(\mathbf{r})\text{d}^2r
\end{eqnarray}
being the expectational value. It is shown that the exciton valley-orbit coupling only
causes energy shifts of degenerate exciton states with angular momentums other than zero.
Since the only difference between the intravalley and intervalley excitons is the exciton
valley-orbit couplings, the binding energies and the wavefunctions of intravalley and
intervalley exctions with zero angular momentum ($1s$ excitons) are the same.

Based on the Hamiltonian given in Eq.~(\ref{intra_exciton_internal_Hamiltonian}), we can solve the exciton energy levels by the variational
method\cite{zhang2019two, wu2019exciton, mypaper0, henriques2021calculation} (see
Appendix~\ref{sec:variation}). In Table.~\ref{tab:TMDCs}, the calculated exciton binding
energies of different intravalley excitons in TMDCs and the parameters used are listed.
The exciton Hamiltonian, screened Coulomb potential, and the calculation method are
given in Appendix~\ref{sec:variation}. It is found that the energy-level splittings
for $2p_{+}/2p_{-}$ excitons are about $10.6$ meV for $\text{MoS}_2$ with $\kappa=1$ and
$10.1$ meV for $\text{MoSe}_2$ with $\kappa=1$. The former value is consistent with the
calculated result in Ref.~\onlinecite{srivastava2015signatures}. However, the latter value
is lower than the experimentally observed value ${14}$ meV in
Ref.~\onlinecite{yong2019valley}. It suggests that the Berry curvature calculated by
the present model might be underestimated in comparison to the full band-structure model.

\subsection{Exciton transitions\label{sec:exciton_transitions}}

\begin{figure}
  \includegraphics[width=0.95\linewidth]{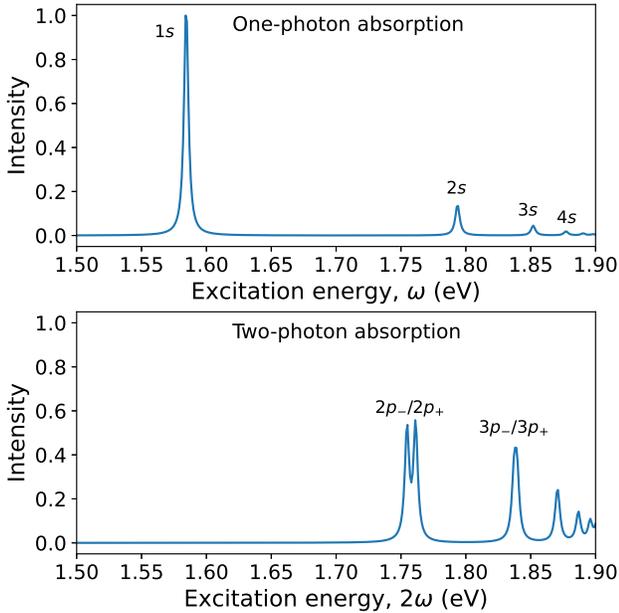}
  \caption{Calculated one-photon (up) and two-photon (down) absorption spectra of
  $\text{MoS}_2$ with the parameters given in Table \ref{tab:TMDCs}., the valley index
  $\tau=1$, and the screening constant $\kappa=2$. The line-broadening factor for the
  two-photon absorption is $\eta=0.1$ eV.}
  \label{fig:optical_spectrum_MoS2}
\end{figure}

In this section, the one-exciton and intra-exciton transitions for intravalley excitons
are studied, and their transition amplitudes are derived. The exciton wavefunction in
momentum space is given by the Fourier transform
\begin{eqnarray}
  \tilde{\Psi}_{\text{X},I\tau}(\mathbf{k})
  &=&
  \int e^{-\mathtt{i}\mathbf{k}\cdot\mathbf{r}}
  \Psi_{\text{X},I\tau}(\mathbf{r})\text{d}^2r
  =
  e^{\mathtt{i}l\varphi_{\mathbf{k}}}\tilde{\mathcal{R}}_{nl}(k),
  \label{exciton_wavefunction1}
\end{eqnarray}
where $\tilde{\mathcal{R}}_{nl}(k)$ is the radial part of the exciton wavefunction in
momentum space, with $n$ the principal quantum number and $l$ the angular momentum. By
using Eq.~(\ref{transformation_of_velocity}) and the FW transformed velocity matrix in
Eq.~(\ref{velocity_matrix}), the momentum matrix element is given by
\begin{eqnarray}
  \tilde{\mathcal{P}}^{\pm}_{\text{eh},\tau}(\mathbf{k})
  &=&
  \left[\tilde{\mathcal{P}}^{\mp}_{\text{he},\tau}(\mathbf{k})\right]^*
  =
  \left[\tilde{\mathcal{P}}^{x}_{\text{eh},\tau}(\mathbf{k})
  \pm\mathtt{i}\tilde{\mathcal{P}}^{y}_{\text{eh},\tau}(\mathbf{k})\right]/\sqrt{2}\n
  &\simeq&
  v_{\text{F}}\left[\delta_{\tau,\pm{1}}\left(1-\frac{\Omega k^2}{4}\right)
  +\delta_{\tau,\mp{1}}\frac{\Omega k^2}{4}
  e^{\pm\mathtt{i}2\varphi_{\mathbf{k}}}\right].\n
  \label{momentum_matrix_element}
\end{eqnarray}
The same result has been derived in Ref.~\onlinecite{gong2017optical}. By inserting the
exciton wavefunction in Eq.~(\ref{exciton_wavefunction1}) and the momentum matrix element
in Eq.~(\ref{momentum_matrix_element}) into the transition amplitude formula in
Eq.~(\ref{transition_amplitude1}), the transition amplitudes for one-exciton transition
can be solved, and the selection rule for linear optical absorption can be derived. It is
found that the absorption of $\mathbf{e}^{+}$ polarized photon generates an exciton at
$\tau=1$ valley with ${l}=0$ or at $\tau=-1$ valley with ${l}=2$, and the absorption of
$\mathbf{e}^{-}$ polarized photon generates an exciton at $\tau=-1$ valley with ${l}=0$ or
at $\tau=1$ with ${l}=-2$, with
$\mathbf{e}^{\pm}=\mathbf{e}_{x}\pm\mathtt{i}\mathbf{e}_{y}$. The selection rule for
one-exciton transition can be summarized as
\begin{eqnarray}
  \begin{cases}
    \Delta{l}+\tau=1 & \text{for $\mathbf{e}^{+}$ polarized photon}\\
    \Delta{l}+\tau=-1 & \text{for $\mathbf{e}^{-}$ polarized photon}
  \end{cases}
\end{eqnarray}
with $\Delta{l}=0,\pm{2}$ and $\tau=\pm{1}$. While the valley index $\tau$ stands for the
angular momentum difference between the ground state and the one-exciton state, the
selection rule can be seen as the consequence of angular momentum conservation in the
photoabsorption process.

To derive the intra-exciton transition, which is defined as the transition between exciton
states\cite{henriques2021calculation}, we consider the exciton internal Hamiltonian with
an in-plane electromagnetic-field interaction. The internal Hamiltonian for intravalley
exctions is written as
\begin{eqnarray}
  \mathcal{H}''_{\text{X},\tau}(\mathbf{r})
  &=&
  \Delta+\frac{p^2}{2\mu}-V(r)-\frac{\Omega}{4}\nabla^2V({r})
  +e\bsym{\mathcal{D}}_{\text{X},\tau}(\mathbf{r})\cdot\bsym{\mathcal{F}}\n
  &&-\frac{\tau\Omega}{2}
  \mathbf{e}_{\perp}\cdot\bsym{\nabla}V({r})\times\mathbf{p},
  \label{internal_hamiltonian2}
\end{eqnarray}
where the exciton dipole-moment operator is defined as
\begin{eqnarray}
  \bsym{\mathcal{D}}_{\text{X},\tau}(\mathbf{r})
  &=&
  \mathbf{r}-\frac{\tau\Omega}{2}\mathbf{e}_{\perp}\times\mathbf{p}.
\end{eqnarray}
The electromagnetic-field interaction can be rewritten as
\begin{eqnarray}
  \bsym{\mathcal{D}}_{\text{X},\tau}(\mathbf{r})\cdot\bsym{\mathcal{F}}
  =
  \mathcal{D}^{+}_{\text{X},\tau}(\mathbf{r})\mathcal{F}^{-}
  +\mathcal{D}^{-}_{\text{X},\tau}(\mathbf{r})\mathcal{F}^{+},
\end{eqnarray}
where $\mathcal{F}^{\pm}=(\mathcal{F}^{x}\pm\mathtt{i}\mathcal{F}^{y})/\sqrt{2}$ and
\begin{eqnarray}
  \mathcal{D}^{\pm}_{\text{X},\tau}(\mathbf{r})
  &=&
  \frac{1}{\sqrt{2}}\left[\mathcal{D}^{x}_{\text{X},\tau}(\mathbf{r})
  \pm\mathtt{i}\mathcal{D}^{y}_{\text{X},\tau}(\mathbf{r})\right]
  =
  r^{\pm}\mp\mathtt{i}\frac{\tau\Omega}{2}p^{\pm},\n
\end{eqnarray}
with $r^{\pm}=(x\pm\mathtt{i}y)/\sqrt{2}=re^{\pm\mathtt{i}\varphi}/\sqrt{2}$ and
\begin{eqnarray}
  p^{\pm}
  =
  -\frac{\mathtt{i}}{\sqrt{2}}\left(\frac{\partial}{\partial{x}}
  \pm\mathtt{i}\frac{\partial}{\partial{y}}\right)
  =
  -\frac{\mathtt{i}}{\sqrt{2}}e^{\pm\mathtt{i}\varphi}\left(\frac{\partial}{\partial{r}}
  \pm\frac{\mathtt{i}}{r}\frac{\partial}{\partial{\varphi}}\right).\n
\end{eqnarray}
Since the exciton dipole-moment operator is related to the dipole-momentum matrix element
by $\mathcal{D}^{\mu}_{\text{X},\tau}(\mathbf{r}) =
\int\tilde{\mathcal{D}}^{\mu}_{\text{X},\tau}(\mathbf{k})
e^{\mathtt{i}\mathbf{k}\cdot\mathbf{r}}{\text{d}^2k}/{(2\pi)^2}$, and the intra-exciton
transition amplitude can be calculated from
Eq.~(28), we find
\begin{eqnarray}
  \xi^{\mu}_{NM}
  =
  -e\int\Psi^*_{\text{X},I_{N}\tau_{N}}(\mathbf{r})
  \mathcal{D}^{\mu}_{\text{X},\tau_{N}}(\mathbf{r})
  \Psi_{\text{X},J_{M}\tau_{M}}(\mathbf{r})\text{d}^2r
\end{eqnarray}
and $|\xi^{\pm}_{NM}|\propto\delta_{l_{N},l_{M}\pm{1}}$. The
selection rule for the intra-exciton transition can be found as
\begin{eqnarray}
  \begin{cases}
    \Delta{l}=1 & \text{for $\mathbf{e}^{+}$ polarized photon}\\
    \Delta{l}=-1 & \text{for $\mathbf{e}^{-}$ polarized photon}
  \end{cases}
\end{eqnarray}
Therefore, given that an $1s$ exciton is generated by the one-exciton transition, an
intra-exciton transition from the $1s$ exciton to $2p_{\pm}$ excitons can be induced by a
two-photon process.

In Fig.~\ref{fig:optical_spectrum_MoS2}, the calculated one-photon (up) and two-photon
(down) absorption spectra of $\text{MoS}_2$ based on the one-exciton transition amplitude
and the intra-exciton transition amplitude are shown. The formula for the one-photon
absorption spectrum is given by Eq.~(\ref{one_photon}) and the formula for the two-photon
absorption spectrum is given by Eq.~(\ref{two_photon}). The resonance peaks in the
one-photon absorption spectrum can be assigned as the transitions of $ns$ excitons, and
the peaks in the one-photon absorption spectrum can be assigned as the transitions of
$np_{-}/np_{+}$ excitons. It is found that the Berry-curvature corrections to the
transition amplitudes are about two order-of-magnitude smaller than the uncorrected
transition amplitudes. Therefore, the contributions of Berry-curvature corrections to the
one-exciton transition probability and the intra-exciton transition probability are not
shown in the optical spectra of TMDCs. The finding is consistent with the calculations  in
Ref.~\onlinecite{gong2017optical}.

\subsection{Exciton valley Zeeman effect}

In this section, the case of an exciton in a out-of-plane magnetic field (${B}\neq{0}$,
$|\mathbf{F}|=|\bsym{\mathcal{F}}|=0$) is considered and the exciton valley Zeeman effect
is studied. Based on the FW transformed single-particle Hamiltonian and the two-particle
potential, the exciton Hamiltonian can be written as
\begin{eqnarray}
  \mathcal{H}_{\text{X}}
  &=&
  \Delta+\frac{|\bsym{\pi}_{\text{e}}|^2}{2m_{\text{e}}}
  +\frac{|\bsym{\pi}_{\text{h}}|^2}{2m_{\text{h}}}-V({r}_{\text{eh}})
  +\frac{\tau_{\text{e}}eB}{2{m}}-\frac{\tau_{\text{h}}eB}{2{m}}\n
  &&
  -\frac{\tau_{\text{e}}\mathbf{e}_{\perp}\cdot\bsym{\nabla}_{\text{e}}V({r}_{\text{eh}})
  \times\bsym{\pi}_{\text{e}}}{4{m}^2v^2_{\text{F}}}
  -\frac{\nabla^2_{\text{e}}V({r}_{\text{eh}})}{8{m}^2v^2_{\text{F}}}\n
  &&
  -\frac{\tau_{\text{h}}\mathbf{e}_{\perp}\cdot\bsym{\nabla}_{\text{h}}V({r}_{\text{eh}})
  \times\bsym{\pi}_{\text{h}}}{4{m}^2v^2_{\text{F}}}
  -\frac{\nabla^2_{\text{h}}V({r}_{\text{eh}})}{8{m}^2v^2_{\text{F}}}.
\end{eqnarray}
Again, we use the coordinate transformation in Eq.~(\ref{relative_coordinate_position})
and Eq.~(\ref{relative_coordinate_momentum}) to rewrite the Hamiltonian by the relative
coordinate system. Additionally, we apply the unitary transformation to the exciton
Hamiltonian, $\tilde{\mathcal{H}}_{\text{X}} =
\mathcal{U}_{\text{X}}\mathcal{H}_{\text{X}}\mathcal{U}^{\dagger}_{\text{X}}$, with the
unitary transformation operator
\begin{eqnarray}
  \mathcal{U}_{\text{X}}
  &=&
  \exp\left[\frac{\mathtt{i}eB}{2}\mathbf{e}_{\perp}
  \cdot\left(\mathbf{R}\times\mathbf{r}\right)\right],
\end{eqnarray}
which causes the momentum operators (note that $\mathbf{p}=-\mathtt{i}\bsym{\nabla}$ and
$\mathbf{P}=-\mathtt{i}\bsym{\nabla}_{\mathbf{R}}$) being transformed as
$\mathcal{U}_{\text{X}}\mathbf{p}\mathcal{U}^{\dagger}_{\text{X}} =
\mathbf{p}-(eB/2)\mathbf{e}_{\perp}\times\mathbf{R}$,
$\mathcal{U}_{\text{X}}\mathbf{P}\mathcal{U}^{\dagger}_{\text{X}} =
\mathbf{P}+(eB/2)\mathbf{e}_{\perp}\times\mathbf{r}$. The exciton Hamiltonian is rewritten
as
\begin{eqnarray}
  \tilde{\mathcal{H}}_{\text{X}}
  =
  \mathcal{H}'_{\text{X}}+\mathcal{H}''_{\text{X}}+\mathcal{J}_{\text{X}},
\end{eqnarray}
where $\mathcal{H}'_{\text{X}}={P^2}/({2m_{\text{X}}})$
%
%
is the exciton translational Hamiltonian,
\begin{eqnarray}
  \mathcal{H}''_{\text{X}}
  &=&
  \Delta+\frac{p^2}{2\mu}-{V}({r})-\frac{\Omega}{4}\nabla^2{V}({r})\n
  &&-\frac{\Omega(\tau_{\text{e}}+\tau_{\text{h}})}{4}
  \mathbf{e}_{\perp}\cdot\bsym{\nabla}{V}({r})\times\mathbf{p}\n
  &&+\frac{eB}{2}\Bigg[
  \left(\frac{1}{m_{\text{e}}}-\frac{1}{m_{\text{h}}}\right)\mathcal{L}(\mathbf{r})
  +\frac{\tau_{\text{e}}-\tau_{\text{h}}}{2\mu}\n
  &&-\frac{\Omega(\tau_{\text{e}}-\tau_{\text{h}})}{4}
  \mathbf{r}\cdot\bsym{\nabla}V({r})\Bigg]+\frac{e^2B^2}{8\mu}r^2
  \label{internal_hamiltonian1}
\end{eqnarray}
is the exciton internal Hamiltonian, and
\begin{eqnarray}
  \mathcal{J}_{\text{X}}
  &=&
  \mathbf{e}_{\perp}\cdot\left[\frac{eB}{m_{\text{X}}}\mathbf{r}
  -\frac{e\Omega(m_{\text{e}}\tau_{\text{e}}-m_{\text{h}}\tau_{\text{h}})}
  {4m_{\text{X}}}\bsym{\nabla}V({r})\right]\times\mathbf{P}\n
  \label{exciton_hamil2}
\end{eqnarray}
is the exciton translational-internal coupling. Note that we have used the identity
$\mathbf{e}_{\perp}\cdot\left[\bsym{\nabla}{V}({r})
\times(\mathbf{e}_{\perp}\times\mathbf{r})\right] =\mathbf{r}\cdot\bsym{\nabla}V({r})$ in
Eq.~(\ref{internal_hamiltonian1}). The first three terms of the exciton internal
Hamiltonian are the Wannier exciton Hamiltonian, the forth term is the Darwin interaction,
the fifth term is the exciton valley-orbit coupling, the sixth term is the valley Zeeman
interaction and the last term is the diamagnetic interaction.

To study the valley Zeeman effect of excitons, it is important to note that the Zeeman
splittings are also contributed from the spin and the atomic orbital of the electron or
hole. The contributions from the Zeeman splittings of spins and atomic orbitals to the
electron energy and the hole energy are given by\cite{bragancca2019magnetic}
\begin{eqnarray}
  \tilde{\varepsilon}_{\text{e}}
  &=&
  \varepsilon^{(0)}_{\text{FW,e}}
  +\frac{s_{\text{e}}g_{\text{spin}}
  +\tau_{\text{e}}g_{\text{e}}}{2}\mu_{\text{B}}B,
\end{eqnarray}
\begin{eqnarray}
  \tilde{\varepsilon}_{\text{h}}
  &=&
  \varepsilon^{(0)}_{\text{FW,h}}
  -\frac{s_{\text{h}}g_{\text{spin}}
  +\tau_{\text{h}}g_{\text{h}}}{2}\mu_{\text{B}}B,
\end{eqnarray}
where $s_{\text{e}}$ and $s_{\text{h}}$ are electron-spin and hole-spin indices,
$g_{\text{spin}}$ is the spin Lande g-factor, $g_{\text{e}}$ and $g_{\text{h}}$ are the
atomic orbital Lande g-factors for electrons and holes, and
$\mu_{\text{B}}={e}/({2m_{0}})$ is the Bohr magneton with $m_0$ the free electron mass.
The band-gap energy is given by
\begin{eqnarray}
  \tilde{\Delta}
  &=&
  \Delta+\frac{1}{2}
  \left[\left(s_{\text{e}}-s_{\text{h}}\right)g_{\text{spin}}
  +\tau_{\text{e}}g_{\text{e}}-\tau_{\text{h}}g_{\text{h}}\right]\mu_{\text{B}}B.
\end{eqnarray}
We assume that the spin and valley indices of the hole are given by
$s_{\text{e}}=s_{\text{h}}=\tau_{\text{h}}=\tau$ for both intravalley and intervalley
excitons, and the valley index of the electron is given by $\tau_{\text{e}}=\tau$ for
intravalley bright excitons and $\tau_{\text{e}}=-\tau$ for intervalley excitons. The
exciton valley Zeeman shift is defined by
\begin{eqnarray}
  E_{\text{X},I} =
  E^{(0)}_{\text{X},I}+\frac{1}{2}g_{\text{X},I}\mu_{\text{B}}B+\cdots,
\end{eqnarray}
where $E^{(0)}_{\text{X},I}$ is the exciton energy level without magnetic-field
interaction, $g_{\text{X},I}$ is the exciton valley g-factor for the $I$-th exciton state.
We can find
\begin{eqnarray}
  g_{\text{X},I}
  &=&
  \left(s_{\text{e}}-s_{\text{h}}\right)g_{\text{spin}}
  +\tau_{\text{e}}g_{\text{e}}-\tau_{\text{h}}g_{\text{h}}\n
  &&+\frac{el_{N}}{\mu_{\text{B}}}
  \left(\frac{1}{m_{\text{e}}}-\frac{1}{m_{\text{h}}}\right)
  +\frac{e(\tau_{\text{e}}-\tau_{\text{h}})}{\mu_{\text{B}}(2\mu)}\n
  &&-\frac{e\Omega(\tau_{\text{e}}-\tau_{\text{h}})}{4\mu_{\text{B}}}
  \langle{\mathbf{r}\cdot\bsym{\nabla}V({r})}\rangle_{I},
  \label{g_factor}
\end{eqnarray}

where the last term in Eq.~(\ref{g_factor}) is an interaction-induced Zeeman splitting.
Assuming $g_{\text{spin}}=2$, $g_{\text{e}}=0$, $g_{\text{h}}=4$ and using the parameters
in the last line of Table.~\ref{tab:TMDCs} ($m_{\text{e}}=m_{\text{h}}=0.44m_{0}$,
$r_0=50.00$, $\kappa=4.0$), the g-factors with the exciton state $I$ assigned as the $1s$
state are given by $|g_{\text{X},1s}|=4$ for the intravalley exciton and
$|g_{\text{X},1s}|=14.5$ for the intervalley exciton, while the interaction-induced Zeeman
splitting contributes about $1.4$ for the g-factor of the intervalley exciton. These
values are consistent with experimental measurements of excitonic states of
$\text{WSe}_{2}$\cite{liu2019magnetophotoluminescence, YCPRL}, where the measured g-factors are
$|g_{\text{X},1s}|\simeq{4}$ for the intravalley exciton and $|g_{\text{X},1s}|\simeq{13}$
for the intervalley exciton.

\section{Summary and perspectives\label{sec:summary}}

In the present work, the FW transformation is applied to 2D gapped Dirac fermions under
external fields. The single-particle Hamiltonian, two-particle interaction, and interband
transition obtained from the FW transformation are used to study the valley-dependent
physical properties of 2D materials. Exciton Hamiltonians for intravalley and intervalley
excitons in an in-plane electromagnetic field and in an out-of-plane magnetic field are
derived. Exciton energy-level splittings, valley-selected exciton transitions, and exciton
valley Zeeman effect are formulated analytically and studied. The variational method is
used to solve the exciton energy levels and optical spectra. The calculated results are
quantitatively coincident with literatures. Even though these effects have been discussed
by different theoretical methods in literatures, we believe that the present theoretical
framework still provides a new viewpoint on these topics and a straightforward derivation
procedure to study related problems.

While the Berry-curvature effects on valley excitons have been studied in literatures,
 many related topics still demand investigation. One
topics of particular importance is the Berry-curvature effect on valley trions and other valley-dependent exciton
complexes. For  problems involving more than two particles, a semiclassical derivation
of Berry-curvature effect becomes much more difficult. Therefore, the present FW
transformation method becomes quite useful for deriving effective Hamiltonians for
valley-dependent exciton complexes. For trions, some theoretical studies addressing the
Berry-curvature effect have been published\cite{hichri2019charged, hichri2020trion}, but
joint effects of valley degree of freedom and external fields on trions have yet to be
studied. Another possible extension of the present framework is to apply the
FW transformation to 2D gapped Dirac fermions with additional interactions or
band-structure modifications. For example, the trigonal warping in the band structures of
2D materials can be included by the FW transformation as a nonlocal kinetic correction to
the exciton Hamiltonian and interband transition. Such extensibility shows the potential
and versatility of the present theoretical framework. These topics will be considered in
future studies.

\section*{acknowledgment}

{This work was supported in part by the Ministry of Science and Technology (MOST), Taiwan
under Contract No. 109-2112-M-001-046 and 110-2112-M-001-042.} Y.-W.C. thanks the
financial support from the Postdoctoral Scholar Program at Academia Sinica, Taiwan, ROC.

\appendix

\section{Variational method\label{sec:variation}}

In the appendix, the variational method to solve the exciton wavefunction and eigenenergy
is introduced. By a rescaling of the length unit to the effective Bohr radius
"$a_0=({m_0}/{m_{\text{e}}})a_{\text{B}}$" and the energy unit to the effective Hartree
"$\varepsilon_{0}=({m_{\text{e}}}/{m_0})(2\text{Ry})$", with $m_0$ the free electron mass,
$a_{\text{B}}\simeq 0.5291772$ \AA\; the Bohr radius and $\text{Ry}\simeq 13.60569$ eV the
Rydberg constant, the exciton Hamiltonian can be rewritten as
\begin{eqnarray}
  \mathcal{H}''_{\text{X}}(\mathbf{r})
  &=&
  -\frac{\sigma+1}{2}\nabla^2-{V}({r})-\frac{\Omega}{4}\nabla^2{V}({r})\n
  &&-\frac{\Omega(\tau_{\text{e}}+\tau_{\text{h}})}{4}
  \frac{1}{r}\frac{\partial{V}(r)}{\partial{r}}\mathcal{L}_{\text{X}}(\mathbf{r}),
\end{eqnarray}
where $\sigma=m_{\text{e}}/m_{\text{h}}$ is the mass ratio, $V(r)$ is a screened Coulomb
potential, $\mathcal{L}_{\text{X}}(\mathbf{r})
=-\mathtt{i}\mathbf{e}_{\perp}\cdot\mathbf{r}\times\bsym{\nabla}
=-\mathtt{i}{\partial}/{\partial\varphi}$ is the angular momentum operator,
\begin{eqnarray}
  \Omega
  =
  \frac{(\sigma+1)^2a^2_0\varepsilon^2_0}{4a^2t^2}
\end{eqnarray}
is the Berry curvature with $a$ the lattice constant and $t$ the hopping coupling. The 2D
screened Coulomb potential is assumed to be given by the Rytova-Keldysh
potential\cite{rytova, Keldysh}
\begin{eqnarray}
  V(r)
  &=&
  \frac{\pi}{\kappa \rho_{0}}\left[H_0\left(\frac{r}{\rho_{0}}\right)
  -Y_0\left(\frac{r}{\rho_{0}}\right)\right]\n
  &=&
  \int\;e^{\mathtt{i}\mathbf{k}\cdot\mathbf{r}}
  \left[\frac{2\pi}{\kappa|\mathbf{k}|\left(1+|\mathbf{k}|\rho_{0}\right)}\right]
  \frac{\text{d}^2k}{(2\pi)^2},
  \label{Rytova_Keldysh}
\end{eqnarray}
where $\rho_{0}=r_0/(\kappa a_0)$ with $r_0$ the screening length in unit of \AA, $\kappa$ is
the dielectric constant, $H_0(r)$ and $Y_0(r)$ are the Struve function and the Bessel
function of the second kind.

We use 2D Slater-type-orbital (STO) as the basis function to expand the exciton
wavefunction\cite{zhang2019two, wu2019exciton, mypaper0}. The exciton wavefunction can be
written as
\begin{eqnarray}
  \Psi_{\text{X},I}
  =
  \sum_{a}c_{a,I}\phi_a(\mathbf{r}),
\end{eqnarray}
where $c_{a,I}$ is the linear variational parameter and
\begin{eqnarray}
  \phi_a(\mathbf{r})
  =
  \frac{e^{\mathtt{i}l_a\varphi}}{\sqrt{2\pi}}r^{n_a-1}e^{-\zeta_a r}
\end{eqnarray}
is the STO with $\zeta_a$ the screening constant and also a variational parameter, $n_a$
the principal quantum number and $l_{a}$ angular momentum. By using linear variational
method, the exciton coefficient can be solved by the eigenvalue equation
\begin{eqnarray}
  \sum_{b}h_{ab}c_{b,I}=\varepsilon_{\text{X},I}\sum_{b}o_{ab}c_{b,I},
  \label{exciton_eigen}
\end{eqnarray}
where $h_{ab}$ is exciton Hamiltonian matrix, $o_{ab}$ is the overlap matrix, and the
eigenvalue of the equation $\varepsilon_{\text{X},M}$ is the exciton energy. The exciton
Hamiltonian matrix is given by $h_{ab} = t_{ab}+v_{ab}+d_{ab}$, with $t_{ab}$ the kinetic
integral, $v_{ab}$ the potential integral, and $d_{ab}$ the band-geometry integral.  The
formulations of the integrals $t_{ab}$, $v_{ab}$, $o_{ab}$ can be found in
Ref.~\onlinecite{mypaper0}. The band-geometry integral is given by
\begin{eqnarray}
  d_{ab}
  &=&
  -\frac{\Omega(\tau_{\text{e}}+\tau_{\text{h}})}{4}
  \int\phi^*_{a}(\mathbf{r})\frac{1}{r}\frac{\partial{V}(r)}{\partial{r}}
  \mathcal{L}_{\text{X}}(\mathbf{r})\phi_{b}(\mathbf{r})\text{d}^2r\n
  &&-\frac{\Omega}{4}
  \int\phi^*_{a}(\mathbf{r})\nabla^2{V}({r})\phi_{b}(\mathbf{r})\text{d}^2r.
\end{eqnarray}

To calculate these orbital integrals containing the screened Coulomb potential, we use the
following formula
\begin{eqnarray}
  \frac{\partial}{\partial{r}}
  =
  \cos\varphi\frac{\partial}{\partial{x}}+\sin\varphi\frac{\partial}{\partial{y}}
  =
  e^{\mathtt{i}\varphi}\frac{\partial}{\partial{z}}
  +e^{-\mathtt{i}\varphi}\frac{\partial}{\partial{z^*}},
\end{eqnarray}
with $z=x+\mathtt{i}y$ and $z^*=x-\mathtt{i}y$, and
\begin{eqnarray}
  \mathbf{k}\cdot\mathbf{r}
  =
  k_{x}x+k_{y}y
  =
  \frac{kz}{2}e^{-\mathtt{i}\varphi_{\mathbf{k}}}
  +
  \frac{kz^*}{2}e^{\mathtt{i}\varphi_{\mathbf{k}}},
\end{eqnarray}
\begin{eqnarray}
  \frac{\partial{V}}{\partial{z}}
  =
  \int e^{-\mathtt{i}\mathbf{k}\cdot\mathbf{r}}
  \left[-\frac{\mathtt{i}k}{2}e^{-\mathtt{i}\varphi_{\mathbf{k}}}\tilde{V}(k)\right]
  \frac{\text{d}^2k}{(2\pi)^2},
\end{eqnarray}
\begin{eqnarray}
  \nabla^2{V}
  =
  \int e^{-\mathtt{i}\mathbf{k}\cdot\mathbf{r}}
  \left[-k^2\tilde{V}(k)\right]\frac{\text{d}^2k}{(2\pi)^2},
\end{eqnarray}
with $\tilde{V}(k) =\int{V}(r)e^{-\mathtt{i}\mathbf{k}\cdot\mathbf{r}} \text{d}^2r$ the
Fourier transform of the screened potential. Given the screened potential being the
Rytova-Keldysh potential, the Laplacian of the potential function can be rewritten by
\begin{eqnarray}
  \nabla^2{V}(\mathbf{r})
  &=&
  -\frac{2\pi}{\kappa \rho_{0}}\int e^{-\mathtt{i}\mathbf{k}\cdot\mathbf{r}}
  \frac{k^2}{k\left(1/\rho_{0}+k\right)}
  \frac{\text{d}^2k}{(2\pi)^2}\n
  &=&
  -\frac{2\pi}{\kappa \rho_{0}}\int e^{-\mathtt{i}\mathbf{k}\cdot\mathbf{r}}
  \frac{\text{d}^2k}{(2\pi)^2}
  +\frac{2\pi}{\kappa \rho_{0}}\int\frac{e^{-\mathtt{i}\mathbf{k}\cdot\mathbf{r}}}
  {1+k\rho_{0}}\frac{\text{d}^2k}{(2\pi)^2}\n
  &=&
  -\frac{2\pi}{\kappa \rho_{0}}\delta(\mathbf{r})
  +\int e^{-\mathtt{i}\mathbf{k}\cdot\mathbf{r}}
  \frac{k}{\rho_{0}}\tilde{V}(k)\frac{\text{d}^2k}{(2\pi)^2}.
\end{eqnarray}
In order to derive the orbital integrals, we can take advantage of the analytic formula
for the Fourier transform of the 2D STO
\begin{eqnarray}
  \tilde{\phi}_{a}(\mathbf{k})
  &\equiv&
  \int\phi_{a}(\mathbf{r})e^{-\mathtt{i}\mathbf{k}\cdot\mathbf{r}}\text{d}^2r
  =
  \frac{e^{\mathtt{i}l_a\varphi_{\mathbf{k}}}}{\sqrt{2\pi}}
  \mathcal{R}_{n_a,l_a}(\zeta_a,k),\n
\end{eqnarray}
where the radial function is generated by\cite{mypaper0}
\begin{eqnarray}
  \mathcal{R}_{n,l}(\zeta,k)
  &=&
  \frac{2\pi(-\mathtt{i})^{n}}{k^{n+1}}\left[\frac{\text{d}^n}{\text{d}z^n}
  \frac{\left(z-\mathtt{i}\eta\sqrt{1-z^2}\right)^{|l|}}{\sqrt{1-z^2}}
  \right]_{z=\mathtt{i}{\zeta}/{k}},\n
  \label{Rk_formula}
\end{eqnarray}
with $\eta=l/|l|$ being the sign of $l$. The band-geometry integral is given by
\begin{eqnarray}
  d_{ab}
  &=&
  \delta_{l_a,l_b}
  \frac{\mathtt{i}l_{b}\Omega(\tau_{\text{e}}+\tau_{\text{h}})}{8}
  \int\Big[\mathcal{R}_{n_a+n_b-2,1}(\zeta_{a}+\zeta_{b},k)\n
  &&+\mathcal{R}_{n_a+n_b-2,-1}(\zeta_{a}+\zeta_{b},k)\Big]\tilde{V}(k)
  \frac{k^2\text{d}k}{(2\pi)^2}\n
  &&-\delta_{l_a,l_b}\frac{\Omega}{4\rho_{0}}\int k\tilde{V}(k)
  \mathcal{R}_{n_a+n_b-1,0}(\zeta_{a}+\zeta_{b},k)\frac{k\text{d}k}{(2\pi)^2}\n
  &&+\delta_{l_a,l_b}\delta_{n_a,1}\delta_{n_b,1}\frac{\Omega}{4\kappa \rho_{0}}.
\end{eqnarray}
The integrals of the transition amplitudes can also be calculated by using the STOs and
the analytical formulation. The one-exciton transition amplitude is given by
\begin{eqnarray}
  j^{\pm}_{N0}
  &=&
  -\frac{ev_{\text{F}}}{(2\pi)^{3/2}}\sum_{\tau,a}c^*_{a,N}\n
  &&\times\Bigg[\delta_{\tau,\pm{1}}\delta_{l_a,0}
  \int^{k_{\Lambda}}_{0}\mathcal{R}^*_{n_a,0}(\zeta_a,k)
  \left(1-\frac{\Omega k^2}{4}\right)kdk\n
  &&+\delta_{\tau,\mp{1}}\delta_{l_a,\pm{2}}
  \int^{k_{\Lambda}}_{0}\mathcal{R}^*_{n_a,\pm{2}}(\zeta_a,k)
  \frac{\Omega k^2}{4}kdk\Bigg]
  \label{one_exciton_transition}
\end{eqnarray}
and $j^{\pm}_{0N}=(j^{\mp}_{N0})^*$. Note that the integration in
Eq.~(\ref{one_exciton_transition}) could diverge as the cut-off momentum
$k_{\Lambda}\rightarrow\infty$. Therefore, the cut-off momentum $k_{\Lambda}=2\pi/a$ with
$a$ the lattice constant is chosen to ensure convergence. The intra-exciton transition
amplitude is given by
\begin{eqnarray}
  \xi^{\pm}_{NM}
  &=&
  -\frac{e}{\sqrt{2}}\sum_{\tau,ab}c^{*}_{a,N}c_{b,M}\delta_{l_a,l_b\pm{1}}
  \frac{(n_a+n_b-1)!}{(\zeta_a+\zeta_b)^{n_a+n_b}}
  \Bigg\{
  \frac{n_a+n_b}{\zeta_a+\zeta_b}\n
  &&\mp
  \frac{\tau\Omega}{2}
  \Bigg[\frac{n_{b}-1\mp l_{b}}{n_a+n_b-1}(\zeta_a+\zeta_b)
  -\zeta_{b}\Bigg]\Bigg\}.
\end{eqnarray}



\end{document}